\documentclass[12pt, onecolumn]{IEEEtran}
\usepackage{algorithm} 
\usepackage{algpseudocode} 
\usepackage{amsmath} 
\usepackage{listings}
\usepackage{algorithm}
\usepackage{graphicx}
\usepackage{subfigure}
\usepackage{float}
\usepackage{lipsum,mwe,cuted}
\usepackage{caption}
\usepackage{subfloat}
\usepackage{amsmath}
\usepackage{cases}
\usepackage{color}
\usepackage{makecell}
\usepackage{amsthm}
\usepackage{cite}
\usepackage{enumerate}
\usepackage{multirow}
\usepackage{makecell}
\usepackage[colorlinks,
linkcolor=black,
anchorcolor=black,
citecolor=black
]{hyperref}
\usepackage{verbatim}

\newtheorem{theorem}{Theorem}

\usepackage[normalem]{ulem}

\usepackage{caption}
\usepackage{mathtools}
\DeclareUnicodeCharacter{2212}{-}
\begin{document}

\title{Performance of Cooperative Detection in Joint Communication-Sensing Vehicular Network: A Data Analytic and Stochastic Geometry Approach}


\author{Hao Ma, Zhiqing Wei, Zening Li, Fan Ning, Xu Chen, Zhiyong Feng\\
	
	\thanks{
		Hao ma, Zhiqing Wei, Fan Ning, Xu Chen and Zhiyong Feng are with
		Key Laboratory of Universal Wireless Communications, Ministry of Education,
		School of Information and Communication Engineering,
		Beijing University of Posts and Telecommunications, Beijing, 100876, China (e-mail: \{steelrickhasen, weizhiqing, fanning, chenxu96330, fengzy\}@bupt.edu.cn).	
		
		Zening Li is with Electrical and Computer Engineering Department, University of Waterloo, Waterloo, Canada, N2L 3G1 (email: z2235li@uwaterloo.ca).}

}

\markboth{}%
{Shell \MakeLowercase{\textit{et al.}}: Bare Demo of IEEEtran.cls for IEEE Journals}

\renewcommand{\thefootnote}{\fnsymbol{footnote}}





\maketitle
\begin{abstract}
The increasing complexity of urban environments introduces additional uncertainty to the deployment of the autonomous vehicular network. A novel road infrastructure cooperative detection model using Joint Communication and Sensing (JCS) technology is proposed in this article to simultaneously achieve high-efficient communication and obstacle detection for urban autonomous vehicles. To suppress the performance fluctuation caused by shadowing and obstruction to the JCS signals, we first derive the statistic of road obstacles from the Geographic Information System (GIS). Then, the analysis of JCS channel characteristics and shadowing factors are presented using Line-of-Sight and Non-Line-of-Sight (LoS and NLoS) channel models under the complex urban scenario. A stochastic geometry approach is applied to analyze the interference factors and the probability distribution of successful JCS detection and communication.
{Simulations have been made to verify the cooperative detection model by probability analysis based on LoS and NLoS channels, and the numerical results demonstrate several different optimization methods for the deployment of JCS road infrastructures. Finally, we simulated and analyzed a deployment optimization method for JCS road infrastructures that complied with the standard of urban traffic-spot structure placement.}
\end{abstract}
\IEEEpeerreviewmaketitle
\section{Introduction}
\subsection{Background and Motivations}
With the acceleration of the global urbanization process, traffic conditions have become much more complex within the urban area. Over 6.5 billion people are estimated to live in the urban area, with more than 2 billion vehicles on the road by 2035\cite{8704212}. Current vehicle manufacturers focus on improving the detection performance of individual vehicle\cite{10.1145/3448076}. However, in the urban area, dense vehicle flow causes severe interference problems to vehicular detection, preventing current Level-2 autonomous vehicles from making safe autopilot decisions\cite{9162145}, making vehicles highly rely on driver control. { The stand of driving automation taxonomy \cite{GBT} has listed the practical requirements to achieve each level of autopilot, and Level-4 and Level-5 require obstacle assistance and networked cooperation to ensure the security of autonomous vehicles and surrounding pedestrians.} {A concept of road infrastructures cooperation\cite{Intel} was proposed to achieve Level-4 autopilot. Whereas the complexity of the urban area dramatically deteriorates the detection performance of infrastructures. Sheltering, signal obstruction, and interference problems\cite{7934405} create blind zones and impose potential risks of accidents on both vehicles and pedestrians.}

To solve the problem above, the joint communication and sensing (JCS) technology has been selected as a potential solution for future vehicular wireless network\cite{9171304}. The JCS road infrastructures use the unified transceiver and spectrum to conduct both communication and sensing simultaneously. Apart from improvement on the efficiency of limited resources in spatial and spectrum dimensions\cite{9359665}, the JCS system enables cooperative detection and coordinated control for multiple road infrastructures\cite{9282206}, which can alleviate the problems of interference. However, obstruction problems within the urban area will lead to severe attenuation for JCS signal and introduce additional attenuation to cooperative detection. {Up to now, the lack of research on infrastructure deployment according to transportation standards has led to additional limits to the solution of the aforementioned problems, such as inter-device interference and blind zones.} {Thus, this paper focuses on the performance analysis of JCS infrastructures using stochastic geometry approaches based on the distribution of obstacles within the urban area by analyzing the measured statistic from the Geographic Information System (GIS), which will ensure the successful cooperative detection and compliance with the standards of road infrastructures in transportation field\cite{STDGAN}.}
\subsection{Related Works}
Early researches on JCS systems mainly focus on waveform design and beamforming techniques\cite{8108565}. {Beyond that, research on the performance of JCS vehicular communication and detection has been proposed by several groups\cite{8382292,9282206,8859331,8453027,9088261,7506244,9171304}.} In \cite{5776640}, a JCS detection method for autonomous vehicles with waveform design and signal processing was proposed. Subsequently, researchers have studied the performance of Pulsed \cite{8461678} and continuous-wave (CW) radar-type-sensor \cite{9411464} based JCS vehicular system, along with waveform optimization and beamforming design\cite{9285278}. An LTE-based V2X network using the JCS technique was presented in \cite{8304814}. {Moreover, a CD-OFDM-based JCS system for machines and vehicles was proposed in\cite{9359665}.}
The conventional deployed Roadside Units (RSUs) are only able to provide relay or communication assistance for vehicular network\cite{8304814}. With the tremendous discrepancy between JCS and RSU devices, RSU cannot achieve the sensing function in its original definition and cannot meet the requirement of JCS cooperative detection\cite{8264740}. A survey about cooperative vehicular networking (CVN) was proposed in \cite{8302837}, introducing infrastructure-to-vehicle (I2V) and infrastructure-to-infrastructure network models and vehicular cooperation methods. {A concept of the vehicular sensor network (VSN) for smart city construction was proposed in \cite{8057297}, introducing traffic management, smart warning, multimedia internet services, reliability of intelligence, urban planning, and environment monitoring technology for intelligent vehicles in the urban area.} Later, a cooperative detection model using JCS-based intelligent road infrastructures was proposed in\cite{9171304}. {Beyond that, a highly efficient information collection and diffusion method for sensing-aided transportation in urban areas was introduced in \cite{8265207}.} Recently, several companies have experimental plans for JCS infrastructures on urban roads\cite{Intel}. 

\begin{figure}[t]
\centering
\includegraphics[width=0.75\linewidth]{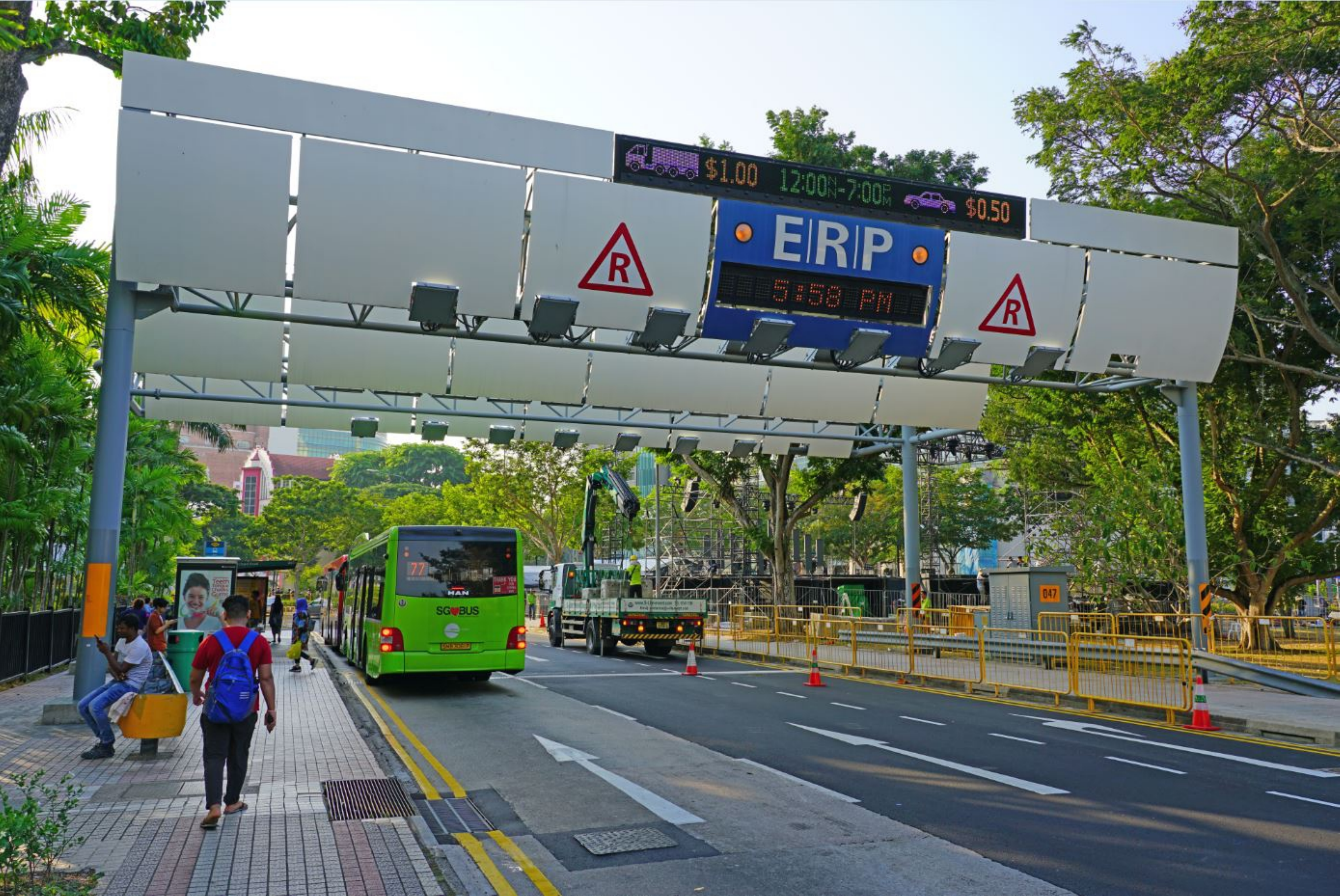}
\caption{Urban Traffic-spot Gantry with Infrastructures\cite{TFGAN}.}
\label{scene1}
\end{figure}

However, the JCS road infrastructures have higher requirements on signal transmission aspect\cite{5948952}, and the deployment optimization approaches for communication base stations\cite{7059538} are not fully complied with JCS networks. {In \cite{7934405}, the authors pointed out that signal shadowing and obstructing will result in significant performance fluctuation in JCS vehicular networks.} JCS signal transmission via the Non-Line-of-Sight (NLoS) channel will suffer severe packet loss. To model the influence of channel fading on the JCS performance, stochastic geometry approaches were proposed to analyze the characteristics of JCS channels\cite{9119440}. In this way, the measured statistic via traffic transportation monitoring is necessary to fully model the distribution of both vehicles and obstacles within the urban area according to the real applications. Besides, the deployment of JCS road infrastructures should also comply with the European installation standard of transportation equipment on cantilever poles, traffic gantry, and other traffic-spot structures\cite{8928298} within the urban area, as shown in Fig. \ref{scene1}. Therefore, the distribution of urban vehicles and obstacles needs to be further analyzed using the measured statistic. Furthermore, a new JCS road infrastructure deployment method compliant with European traffic standards is required, and performance analysis of JCS cooperative detection is required to validate the efficiency of JCS road infrastructure.
\subsection{Our Contributions}
This paper proposes a novel JCS infrastructure cooperative detection model for I2V cooperation.
{The measured statistic of vehicles and obstacles is analyzed to study the characteristics of LoS and NLoS channels in the urban area.}
{The JCS detection and communication performance are analyzed based on the stochastic geometry approach by calculating correlated scattered-signals and inter-device interference under LoS and NLoS channels. }
An algorithm to derive the numerical results of JCS cooperative detection performance is then proposed.
Moreover, according to transportation standards\cite{STDGAN}, a deployment method for infrastructures is proposed to optimize the performance of JCS cooperative detection. 
The main contributions of this paper are summarized as follows:
\begin{enumerate}
	\item We propose a novel JCS cooperative detection model compatible with European urban transportation standards for traffic-spot infrastructures in the urban area.
	\item We analyze a set of measured statistic from the GIS.
	The distribution model of obstacles within the urban area is analyzed. The LoS and NLoS channel model is established based on the analysis of obstacle distributions.
	\item We calculate the probability of successful JCS detection and communication based on the LoS and NLoS channel model. 
	Further, we propose the numerical analysis of JCS cooperative detection performance within an algorithm.
	\item We analyze the probability of successful cooperative detection and the coverage probability of infrastructures. Further, a deployment adjusting method is proposed to optimize the performance JCS cooperative detection.
\end{enumerate}
\begin{figure}[t]
\centering
\includegraphics[width=0.85\linewidth]{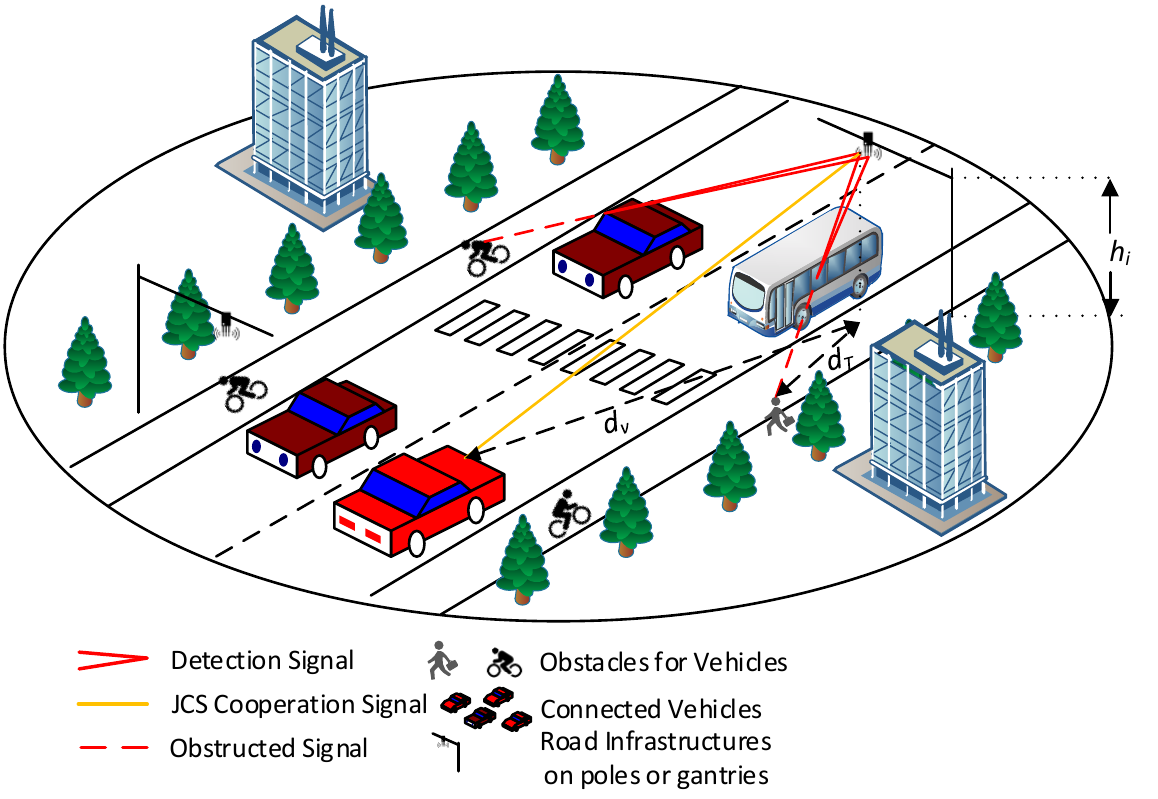}
\caption{Urban infrastructures network model.}
\label{scene01}
\end{figure}
\begin{figure*}[!t]
	\normalsize
	\centering
	\includegraphics[width=0.9\linewidth]{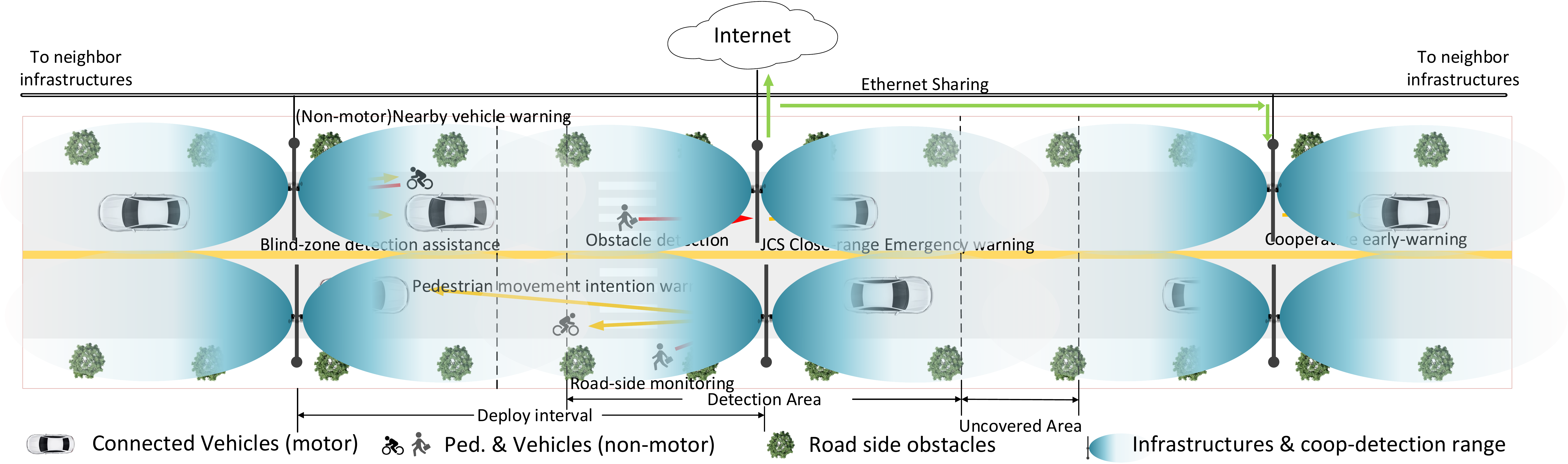}
	\caption{Cooperative detection model for road infrastructures.}
	\hrulefill
	\label{Coopmodel}
	\vspace*{4pt}
\end{figure*}
\subsection{Outline of This Paper}
{The remaining parts of this article are organized as follows. In Section \ref{SECII}, we propose the system model and analysis of GIS statistics for obstacle distributions. In Section \ref{SECIII}, the performance analysis of JCS infrastructures cooperative detection using the stochastic geometry approach is introduced. Section \ref{SECIV} gives the numerical and simulation results of JCS infrastructures' cooperative detection performance. Section \ref{SECV} concludes this article.}

\textbf{Notations:} The summary of notation is shown in Table \ref{notation}

\begin{table}[t]
	\centering
	\caption{Summary of notification} 
	\begin{tabular}{m{2.0cm}<{\centering}|m{5.4cm}<{\centering}}
		\hline
		\hline 
		\textbf{Notation} 	& \textbf{Meaning/Defination} 	\\ 
		\hline 
		$h$			& Height of infrastructures.	\\ 
		$d$ 		& Distance between infrastructure and objects.	\\
		$d_T$		& Distance between infrastructure and obstacles.	\\ 
		$d_v$		&	Distance between infrastructure and vehicles.	\\
		$P_t$ 		& 	Base-band transmitting power.\\
		$G_t$ 		& 	Transmitter antenna gain.\\
		$G_r$		&	Receiver infrastructure antenna gain.\\
		$G_{rc}$		&	Receiver communication antenna gain.\\
		${\rm Pr}_{\rm LoS}(d, h_1, h_2)$& Probability of LoS transmission.\\
		$h_1$		&	Height of transmitter.\\
		$h_2$		&	Height of receiver.\\
		$|x|$		&	Distance between the transmitter and receiver.\\
		$|x_i|$		&	The distance from $i$-th interfering device.\\
		$\eta$		&	Attenuation factor for the NLoS propagation.\\
		$P_{rc}$	&	Power of received communication signal.\\
		$\alpha$	&	Channel loss exponent.\\ 
		\hline 
		\hline 
	\end{tabular} 
	\label{notation} 
\end{table} 
\section{System Model}\label{SECII}
We consider a JCS infrastructure network that conducts cooperative detection to sense the road environment. The deployment of the JCS infrastructure complies with the transportation standards \cite{ITU-TK20}. The JCS infrastructure network model, cooperative detection model, and road obstacle statistic and distribution Model are provided in detail as follows.
\subsection{JCS Infrastructure Network Model}
{According to the road infrastructure deployment standard released by ITU-T \cite{ITU-TK20}, installation points on gantries and cantilever poles are recommended as the deploy position of road infrastructures.} 
{The JCS infrastructure network model is shown in Fig. \ref{scene01}. The Infrastructures monitor obstacles and vehicles on the urban road and roadside area by conducting JCS detection with multi-beam to track multiple targets on the road. The fusion centers integrate monitoring data from multiple infrastructures and send it to vehicles on the road via the communication function of JCS infrastructures. }
However, signal obstruction occurs frequently within the complex urban area, distressing the cooperation of infrastructures. 
\subsection{Cooperative Detection Model}\label{SMCOOP}
{The infrastructures cooperative detection model consists of infrastructure-to-obstacle (I2O) detection, I2V tracking, and data sharing, illustrated in Fig. \ref{Coopmodel}. Three typical types of JCS infrastructure cooperation are designed as follows: }

\begin{enumerate}
\item The JCS infrastructures perform feedback-based JCS local cooperation to track and communicate with the vehicles simultaneously. 
\item Emergency information could be transported to neighbor JCS infrastructures for early warning.
\item The JCS infrastructures perform I2O detection in close-range coverage to sense and track obstacles, including road-area and roadside monitoring, non-motors/pedestrians tracking, and emergency detection. {After that, the data of I2O detection is shared with vehicles for JCS I2V cooperative detection.}
\end{enumerate}

\begin{figure}[t]
	\centering
	\includegraphics[width=1\linewidth]{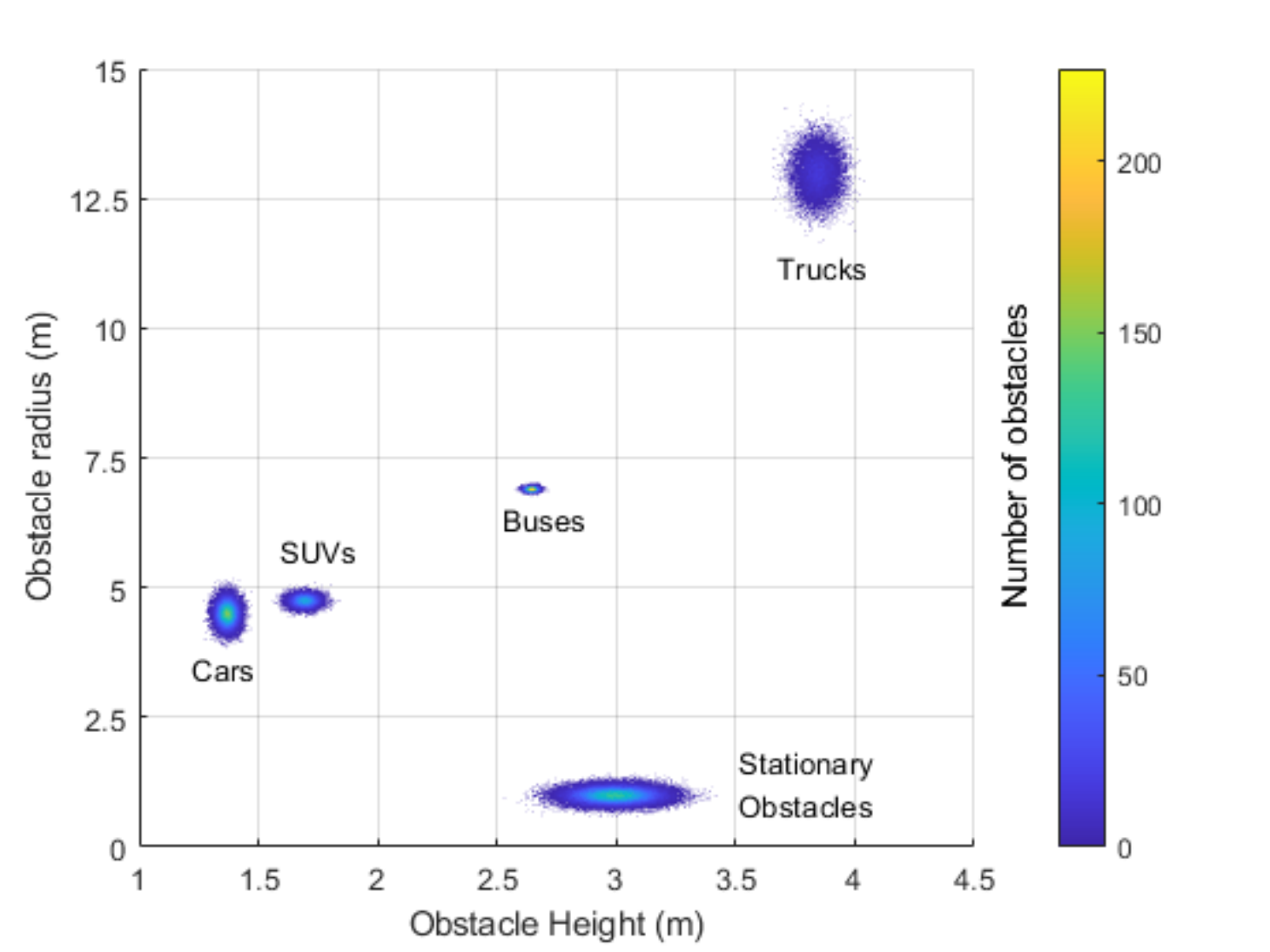}
	\caption{Distribution of stationary and mobile obstacles.}
	\label{OBS2DS}
\end{figure}

\begin{figure*}[h]
	\normalsize
	\centering
	\centering
	\subfigure[Statistic fitted by measured height of obstacles and vehicles]
	{
		\includegraphics[width=0.20\linewidth]{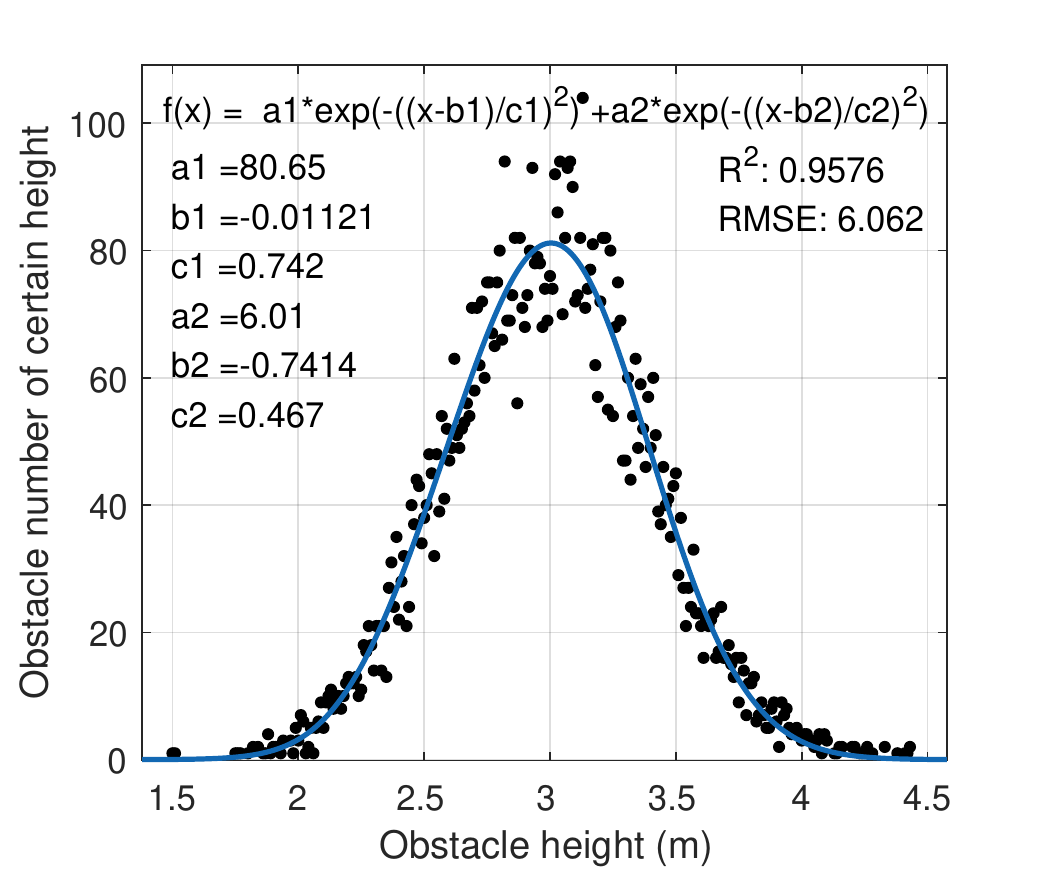}
		\label{data1-1}
		\includegraphics[width=0.20\linewidth]{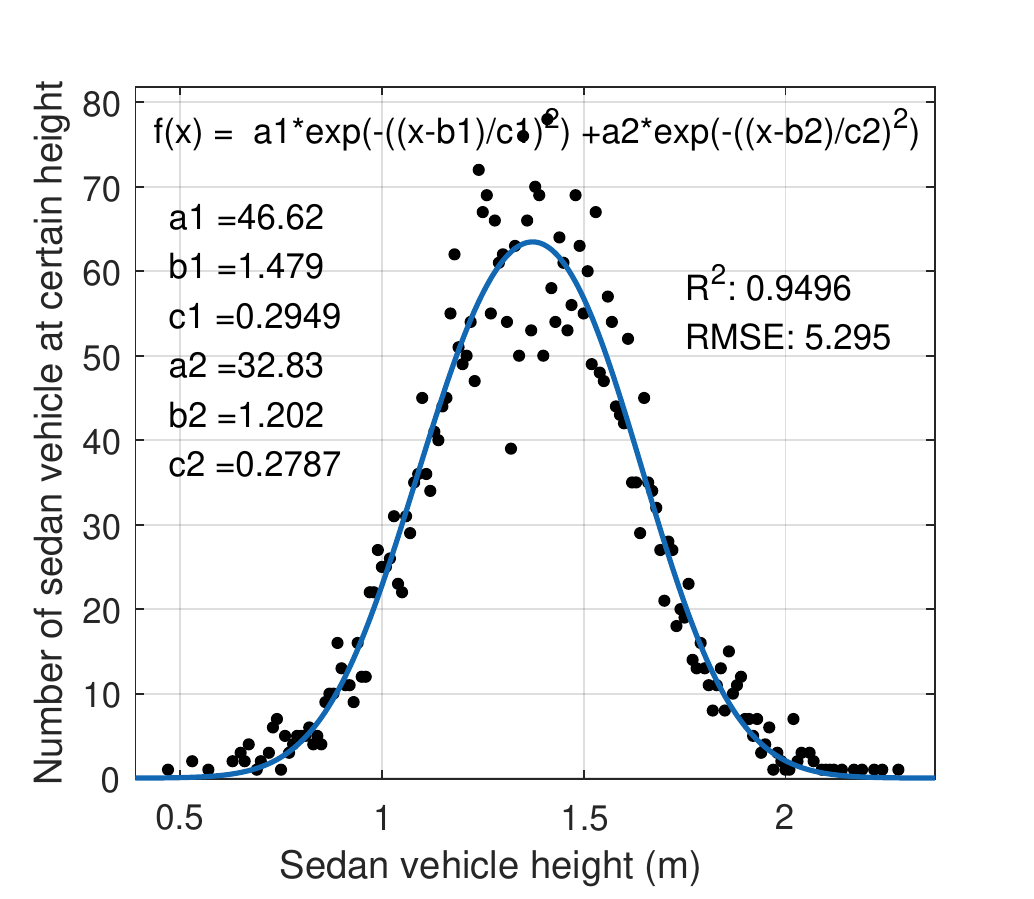}
		\label{data1-2}
		\includegraphics[width=0.20\linewidth]{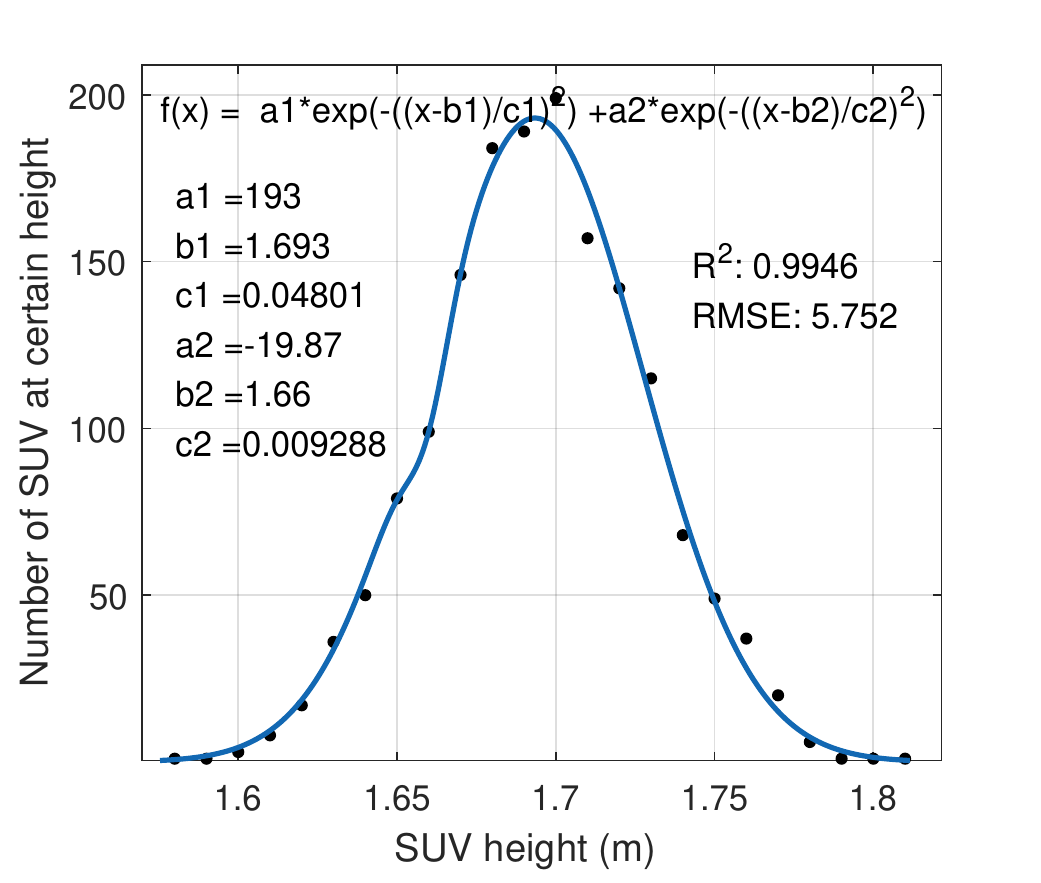}
		\label{data1-3}
		\includegraphics[width=0.20\linewidth]{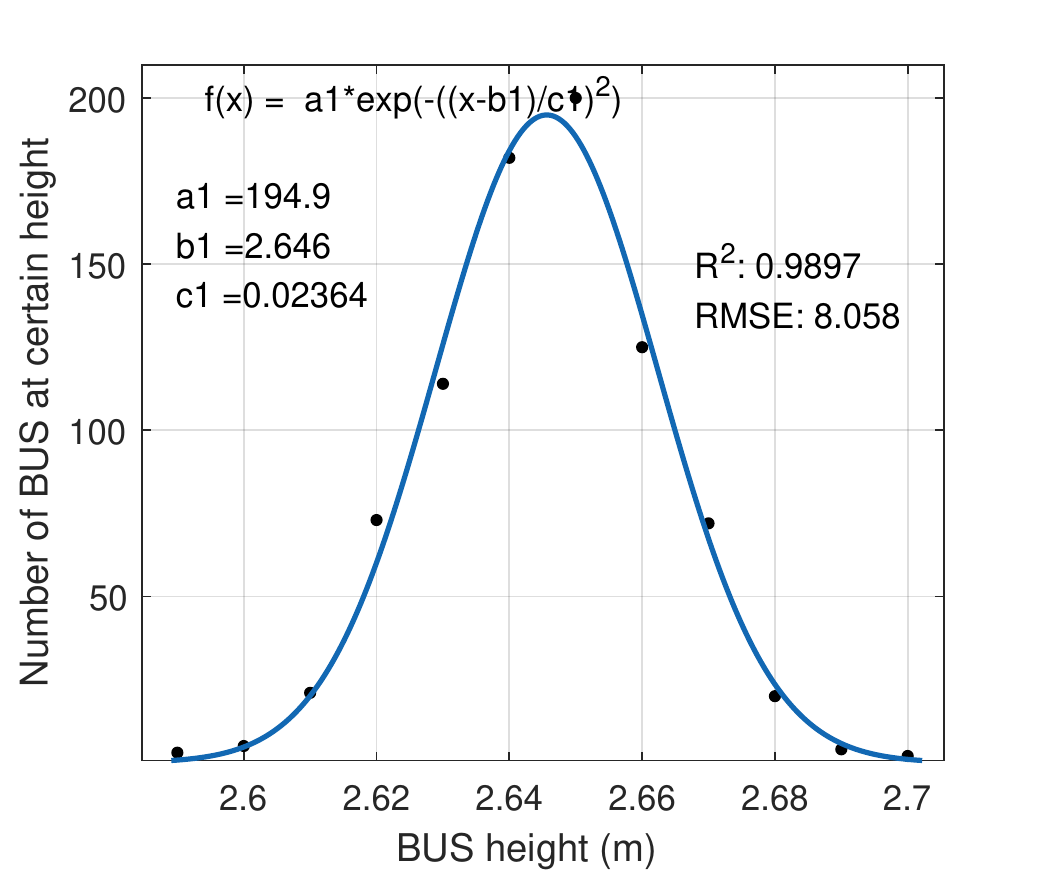}
		\label{data1-4}
		\includegraphics[width=0.20\linewidth]{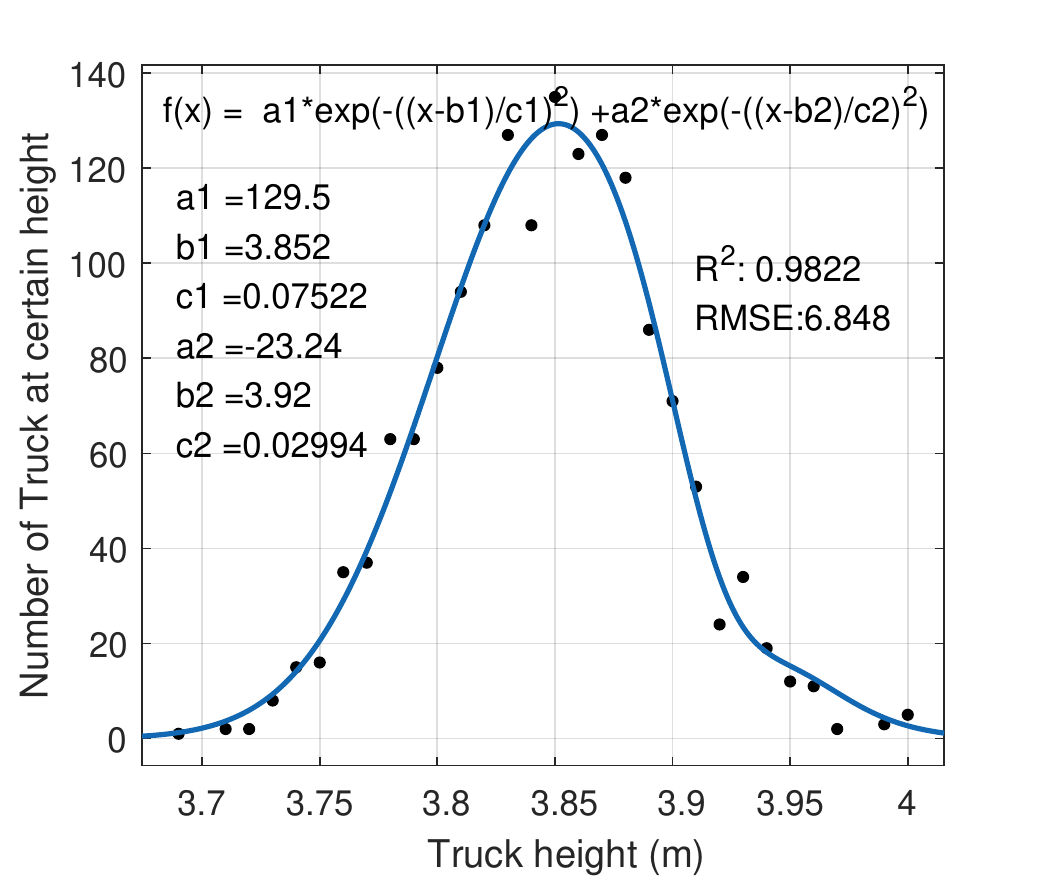}
		\label{data1-5}
	}
	\\
	\subfigure[Statistic fitted by measured radius of obstacles and vehicles]
	{
		\includegraphics[width=0.20\linewidth]{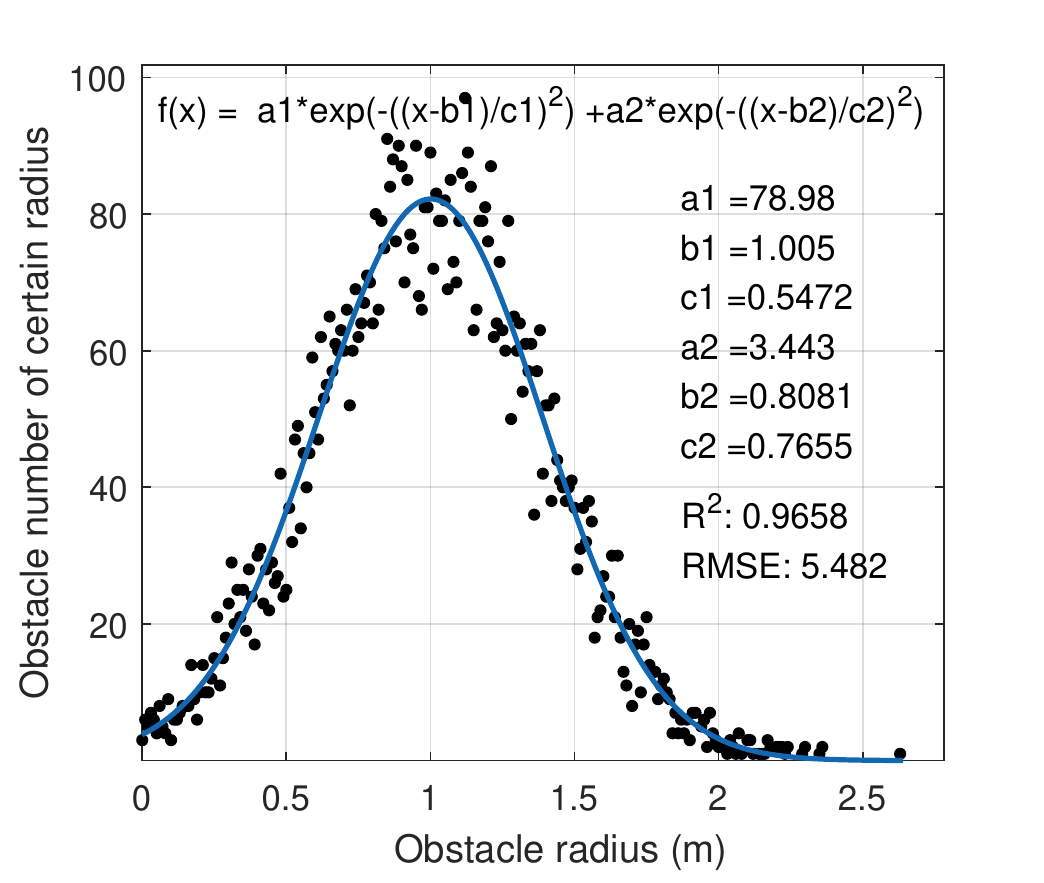}
		\label{data2-1}
		\includegraphics[width=0.20\linewidth]{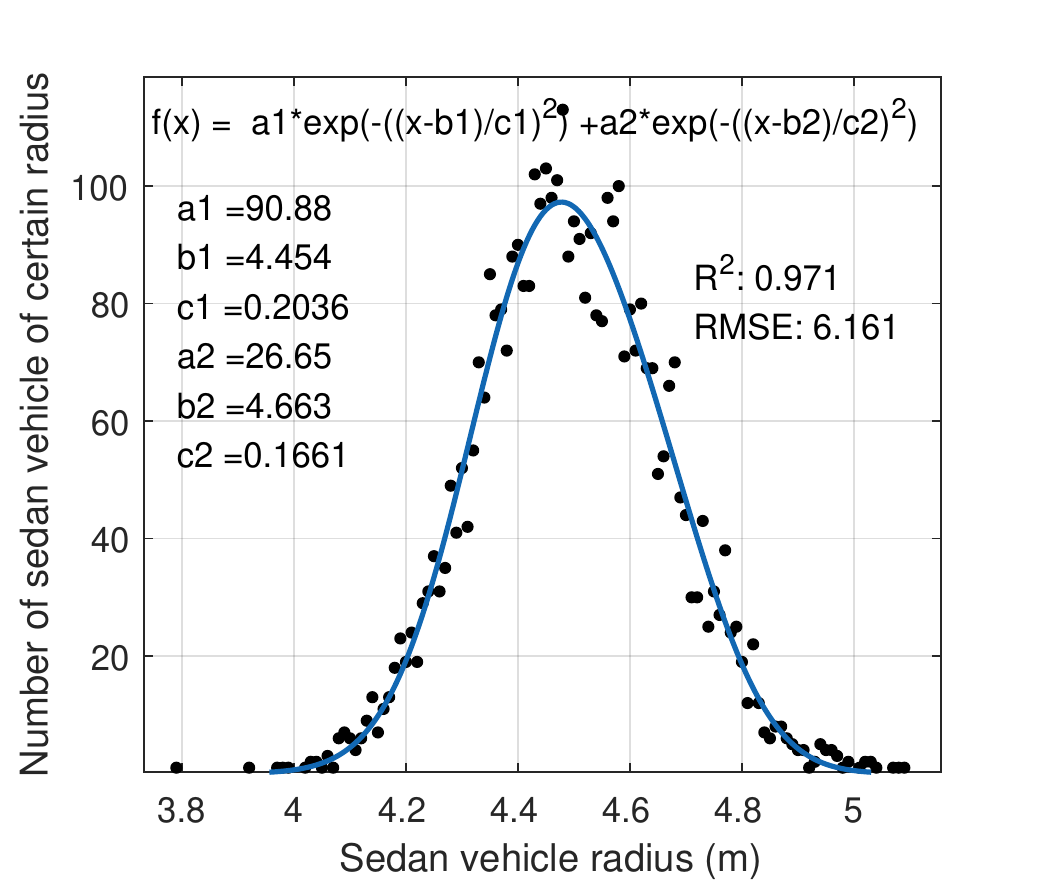}
		\label{data2-2}
		\includegraphics[width=0.20\linewidth]{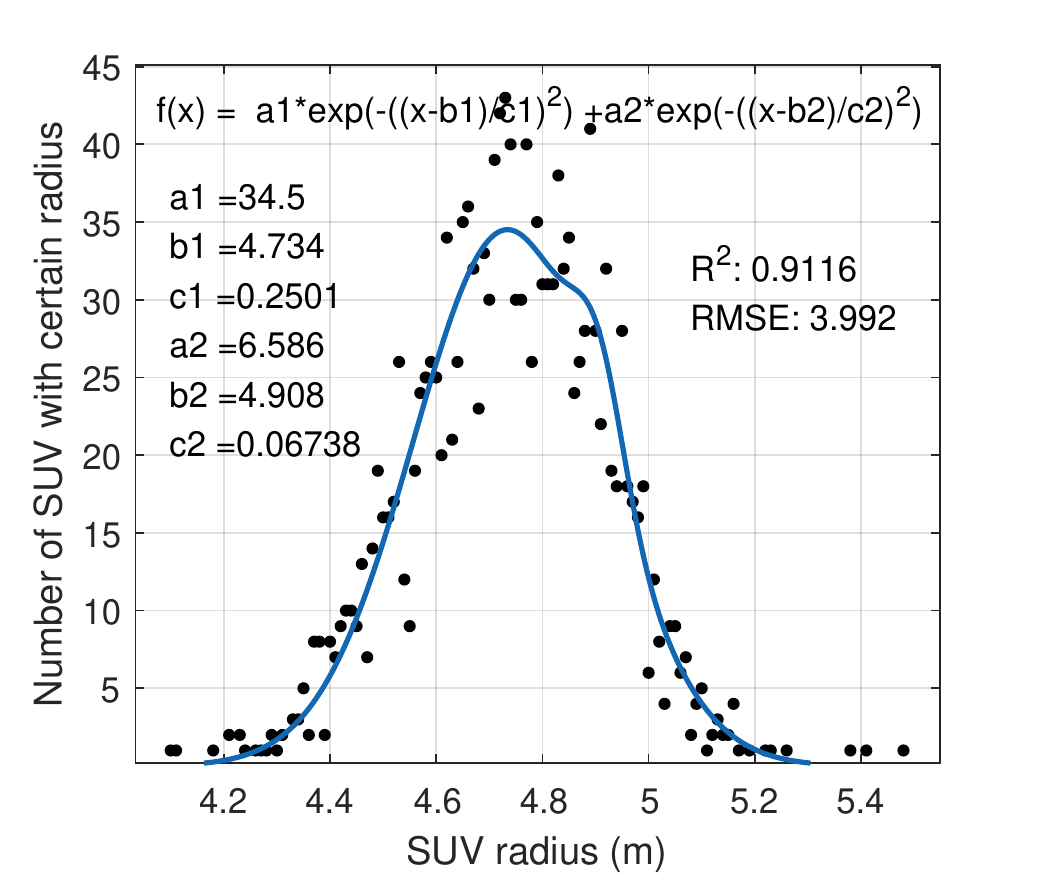}
		\label{data2-3}
		\includegraphics[width=0.20\linewidth]{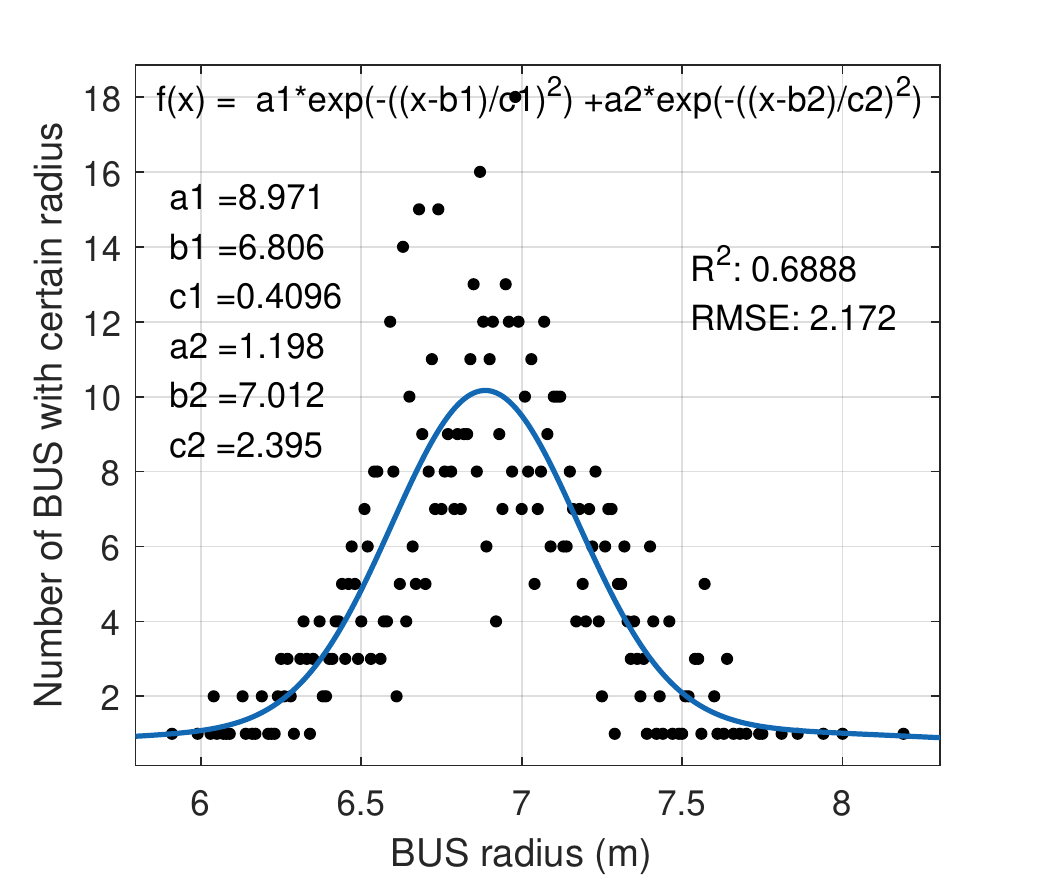}
		\label{data2-4}
		\includegraphics[width=0.20\linewidth]{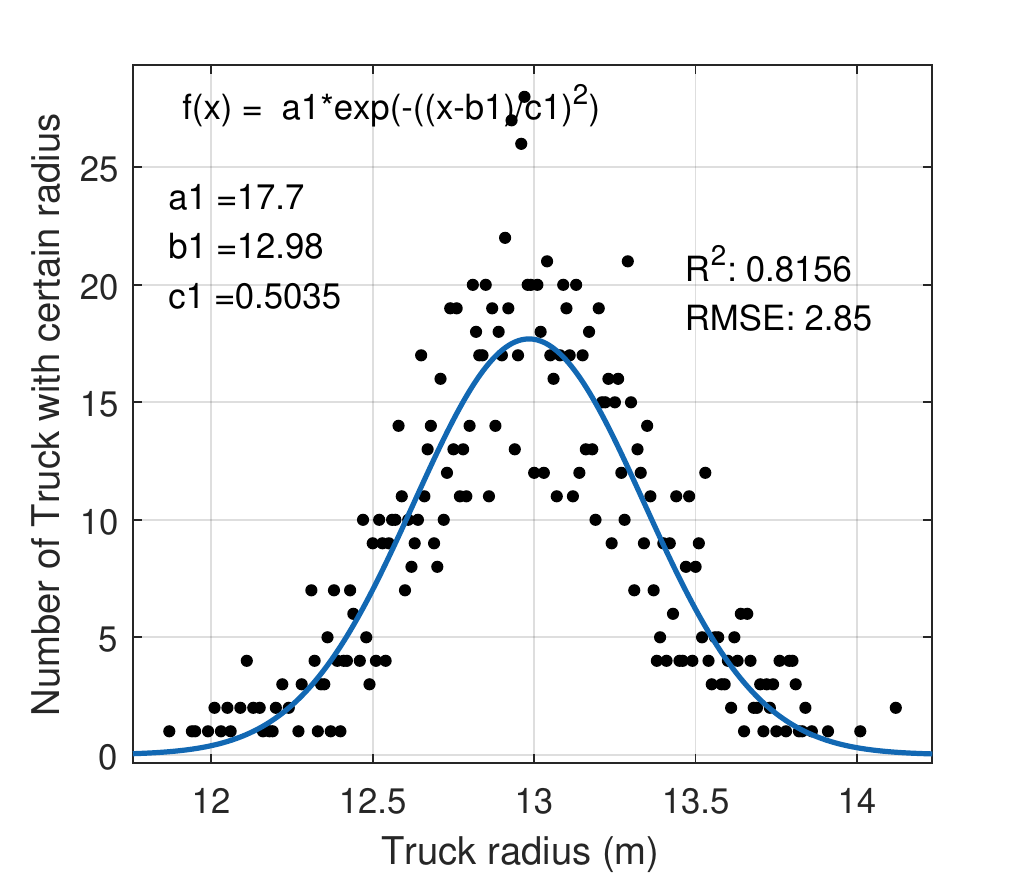}
		\label{data2-5}
	}
	\caption{Statistic and fittings of measured data for obstacles and vehicles within urban area.}
	\hrulefill
	\label{data}
	\vspace*{4pt}
\end{figure*}

Successful I2O, I2V detection and I2V data sharing are necessary to ensure successful I2V cooperative detection.
{However, dense vehicles and obstacle distribution within the urban area lead to a severe dynamic obstruction to JCS signals, introducing fluctuation to LoS transmission and successful detection probability, which introduces an additional risk of accidents to autonomous vehicles.} The adjustment of JCS infrastructure deployment can ameliorate the problem of signal obstruction and increase LoS transmission probability.
\subsection{Obstacle Statistic and Distribution Model}\label{ObsANA}

\begin{figure}[t]
	\centering
	\includegraphics[width=1\linewidth]{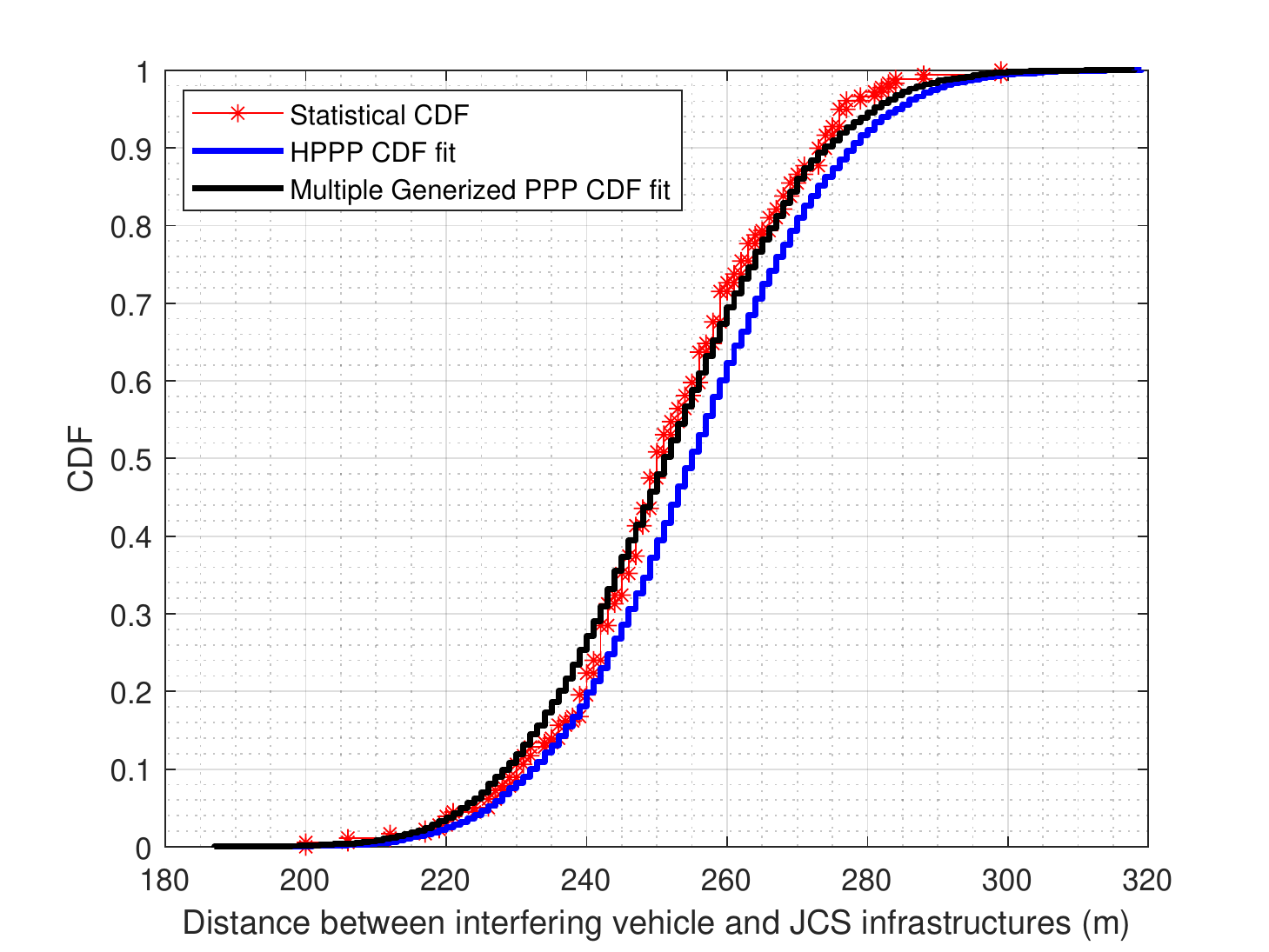}
	\caption{Distribution of stationary and mobile obstacles.}
	\label{IDis}
\end{figure}

Both vehicles and stationary obstacles can obstruct JCS signals, causing JCS cooperative detection performance fluctuation.
In this paper, we utilize vehicular and obstacle data from Geographical Information Monitoring Cloud (GIM cloud) Platform\cite{GIM_cloud} to characterize the distribution of multiple types of vehicles and stationary obstacles. {The selected dataset in this paper consists of virtual map statistics such as vector data of transportation, communication, navigation and positioning, which is collected through urban geo-surveying services, optical recording and other geographical measurement approaches.
The number of vehicles and obstacles are recorded along with statistics of height and radius on the projection plane. Measured statistics of objects' height and radius are classified by stationary obstacles and vehicle as sedans, SUV, buses and trucks.}  
{For each type of vehicle or obstacle, the Gaussian Mixture Model is applied to fit the distribution of vehicle and obstacle height and radius. The result of fitting is derived as $f(x)=\sum_{i=1}^{n}A_i e^{-(x-B_i)^2/2C_i^2}$ with fitting parameters $A_i$, $B_i$ and $C_i$ and the fitting results shown in Fig. \ref{data}. Besides, TABLE \ref{Dis_Tab} summarize the value of fitting parameters for GMM models in Fig. \ref{data}. 
The parameters are used to build the LoS and NLoS channel models and the performance of JCS cooperative detection in the subsequent sections.
\begin{table}[t]
	\centering
	\caption{Parameters of Fitting Function} 
	\begin{tabular*}{8cm}{lllll} 
		\hline 
		Obstacle 	& Height & Height 		& Radius 	& Radius 	\\ 
		&mean (m)& std (m)	& mean (m)	&std (m)	\\ 
		\hline 
		Sedans 	& 1.3750 & 0.0250 		& 4.5001 	& 0.1657 	\\ 
		SUVs 		& 1.6998 & 0.0336 		& 4.7494 	& 0.0829 	\\
		Buses		& 2.6498 & 0.0169 		& 6.8995 	& 0.0333 	\\ 
		Trucks 		& 2.5670 & 1.8144 		& 8.6676	& 6.1307	\\
		Stationary 	& 2.9951 & 0.1003		& 0.9953	& 0.1000	\\
		\hline 
	\end{tabular*} 
	\label{Dis_Tab} 
\end{table} 

Beyond that, the statistic of the infrastructures to vehicles distance $d_v$ is also analyzed to summarize the distribution of vehicle and obstacle distance. The distributions of height and radius are presented in Fig. \ref{OBS2DS}, and the cumulative distribution function of $d_v$ is shown in Fig. \ref{IDis}. As the vehicles and obstacles are constructed by standard manufacturing procedures in the factories, we use the discrete Gaussian model to fit with the distribution of the height and radius for vehicles and obstacles. Different from the distribution of height and radius, the position of each vehicle and obstacle should be randomly distributed on the road. The Poisson Point Process is always considered a suitable model to fit with the random distribution of a set of points. However, vehicles usually have a random protection interval and cannot overlap with other vehicles or obstacles, which is different from other scenarios that could ignore the overlapping problem of points. In this way, the Poisson Point Process model could not truly depict the distribution model of vehicles and obstacles, especially in the complex urban area. Based on the statistics of measured infrastructure to vehicle distance, a Superimposed Generalized PPP to fit with the distribution of infrastructure to vehicle distance. As shown in Fig. \ref{IDis}, the measured distribution of vehicles follows a convergence of 3 different distributions, which could be expressed by:}
{
	\begin{align}
	X(d) = X_1({d})+X_2({d})-oX_{\tau}(d'),
	\end{align}
}
{where $X_1(d)$ and $X_2(d)$ is independent finite point sets both following PPP with $\lambda_{X1} = 150$, $\lambda_{X2} = 5$, $oX_\tau(d')$ is a PPP with $\lambda_{X0} = -3.65$ to adjust and randomly remove the points which is within the minimum protection interval of another point, which may have a different distribution and the number of points from other processes.}

\subsection{Communication Channel Model}\label{sbusectionCommchannel}
The transmitting power of a single JCS beam is denoted by $P_t$. Then, the received power of I2V transmission is given by
\begin{align} \label{equ1}
	{{P}_{rc}}{=}\left\{ \begin{array} {lcl}
	{{P}_{t}}{{\left| x \right|}^{-{{\alpha }}}} & & \rm LoS \\ 
	\eta {{P}_{t}}{{\left| x \right|}^{-{{\alpha }}}} & & \rm NLoS \\
	\end{array}, \right.\
\end{align}
where $\alpha$ is the free-space channel loss exponent, $\eta$ is the attenuation factor for the NLoS propagation, and $|x|$ is the distance between the transmitter and receiver. Thus, ${{P}_{rc}}$ is derived as
\begin{align}
&{{P}_{rc}} ={{P}_{rc,\rm LoS}}+{{P}_{rc,\rm NLoS}} \nonumber\\ 
& ={{P}_{t}}{{\left| x \right|}^{-{{\alpha }}}}{{Pr }_{\rm LoS}}+\eta {{P}_{t}}{{\left| x \right|}^{-{{\alpha }}}}(1-{{Pr }_{\rm LoS}}),
\end{align}	
where ${{P}_{ rc,\rm LoS}}$ and ${{P}_{rc,\rm NLoS}}$ are power of communication signals via LoS and NLoS channels, respectively.
The directional communication signals transmitted via the NLoS channel are obstructed by obstacles and vehicles in the urban area. Thus, the probability of LoS transmission is decided by obstacle distributions, which is expressed as\cite{9052010}
\begin{align}\label{equ2}
	{{Pr}_{\rm LoS}}(d,{{h}_{1}},{{h}_{2}})=\exp (-2{{r}_{0}}{{\lambda }_{0}}\int_{0}^{d-\frac{\pi }{2}{{r}_{0}}}{G(h)dx}),
\end{align}
\begin{figure}[t]
	\centering
	\includegraphics[width=1\linewidth]{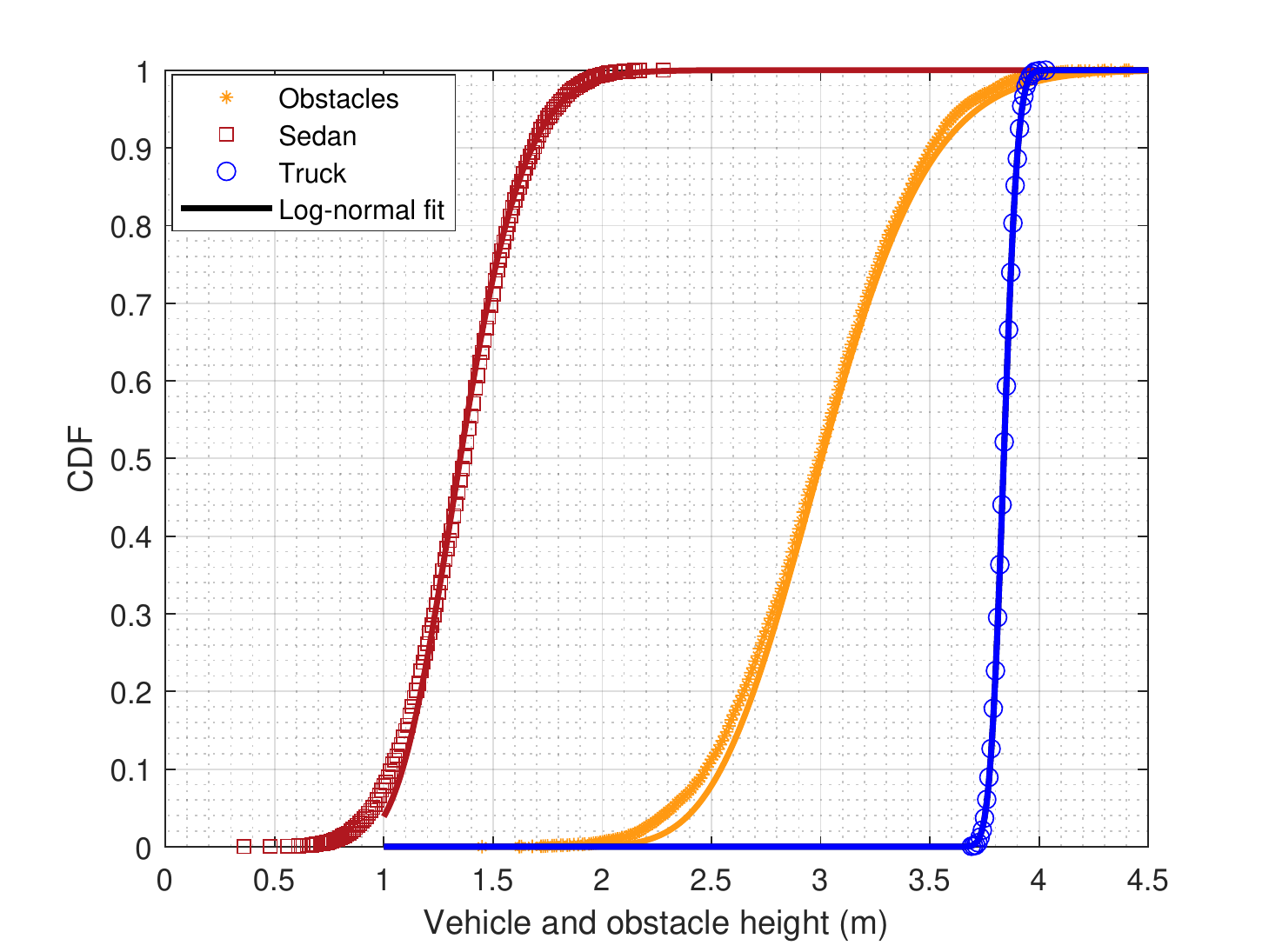}
	\caption{The height distribution of the typical vehicles and obstacles, compared by the log-normal and the empirical data. [$\mu_0$, $\sigma_{0}$] of the typical data are [1.1, 0.13], [0.3, 0.17], and [1.345, 0.013] for obstacles, sedans, and trucks respectively.}
	\label{CDFH}
\end{figure}
{where the equivalent radius of vehicles or obstacles is $r_0$, the density of the obstacles and vehicles is $\lambda_{0}$, and $G(h) = 1- F(h)$ is the complementary cumulative distribution function (CDF) of the real-time measured data from the GIS platform. Fitted with the suggested log-normal model based on \cite{9052010}, the empirical CDF of the GIS data shown in Fig. \ref{CDFH} can be expressed as:}
\begin{align}
F(h)=\frac{1}{2}+\frac{1}{2}erf\left[ \frac{\ln h-{{\mu }_{0}}}{\sqrt{2}{{\sigma }_{0}}} \right],
\end{align}
where $erf(\cdot)$ is the error function, and $\mu$ and $\sigma_0$ are
\begin{align}
	{{\mu }_{0}}=\ln \left[ \frac{{{m}_{0}}}{\sqrt{1+\frac{{{\nu }_{0}}}{m_{0}^{2}}}} \right], \sigma _{0}^{2}=\ln \left[ 1+\frac{{{\nu }_{0}}}{m_{0}^{2}} \right],
\end{align}
respectively, with $m_0$ and $\nu_0$ being the mean and various of the height for obstacles, which are obtained from the parameters in Section \ref{ObsANA}.

The probability of successful communication, denoted by $\rm Pr_{succ,c}$, is decided by the power of the received communication signal $P_{rc}$ and interference factors. The communication receiver has a threshold $\beta_c$, and the signal to noise plus interference ratio (SINR) of the received JCS signal should be higher than the threshold to ensure successful communication. The $\beta_{c}$ on vehicles and machines using JCS devices are 7 dB considering the limit of NI JCS antennas\cite{9359665}. Further analysis of $\rm Pr_{succ,c}$ is shown in Section \ref{SECIAC}.
\subsection{Infrastructure Detection Model}\label{SubsecID}
JCS infrastructures rely on the echoes of JCS signals to detect vehicles and obstacles. The channel model decides the fading of JCS echoes, which determines the successful target sensing, tracking, and range and velocity measurements. {The JCS infrastructures detect targets with $n_p$ beams for a single round of detection\cite{8855937}.} In the urban area, each beam is transmitted via an independent LoS or NLoS channel with random signal obstruction. The numbers of echoes via LoS and NLoS channels are denoted by $n_{\rm ps}$ and $n_{\rm ps,NLoS}$ respectively, and are expressed as
\begin{align}
n_{ps} &= n_p {{\rm Pr }_{\rm LoS}}(d,{{h}_{1}},{{h}_{2}}), \rm and\\
n_{ps,\rm NLoS} &= n_p (1-{{\rm Pr }_{\rm LoS}}(d,{{h}_{1}},{{h}_{2}})).
\end{align}
{The probability of LoS transmission indicates the multi-path characteristics of the received signals. Received signals from multi-oath echoes will be considered recoverable only when the signal-to-noise-and-interference-ratio (SINR) meets the requirements of JCS antennas. If the multi-path signals are recoverable, then JCS infrastructures could perform NLoS detection to obtain data on over-the-horizon targets. However, signals from either multi-path or LoS paths that do not meet the requirement of the SINR threshold will be considered as scattered and added to the interfering factors. Under this situation, the channel of JCS detection and communication is sparse \cite{9373010}.}
Unlike the additive characteristics of communication channels, each JCS sensing beam has an independent successful detection probability. The overall probability of successful detection depends on the number of effective beams that successfully detect the target. The JCS receivers also have an SINR threshold for sensing denoted by $\beta_{s}$, which is 13 dB with NI antenna and signal processing devices for JCS road infrastructures\cite{9359665}. 
\section{Deployment and Performance analysis of JCS Road Infrastructures Cooperative Detection}\label{SECIII}
As demonstrated in Section \ref{SMCOOP}, the cooperative detection model of JCS infrastructures consists of I2O, I2V detection, and I2V communication. Performance of sensing and communication determines the JCS cooperative detection performance\cite{8253543}. This section presents the performance analysis of sensing and communication under the urban LoS and NLoS channel models. The probability of successful cooperative detection is derived by analyzing the joint probability distribution of I2O, I2V detection, and I2V communication. The algorithm to calculate the probability of coverage, the probability of effective detection and communication, and the JCS cooperative detection range is proposed.
\subsection{Performance of JCS Infrastructure Detection} \label{detectionsignal}
The performance of JCS detection is evaluated by the successful JCS detection probability. Decisive metrics of JCS detection performance consist of attenuation characteristics and interference of JCS echoes. The performance of infrastructure detection under the urban LoS and NLoS channel model is analyzed as follows.
\subsubsection{Attenuation Analysis of JCS Detection Echoes}
{The power of echoes for a single JCS beam is denoted by $P_{res}$. Based on\cite{Radar-Handbook}, we transform the radar range equation to calculate the $P_{res}$. Since the signal is echoed from the target with radar cross section (RCS) as $S_{ref}$ at the distance $r$, The original form of radar range equation is 	 
	\begin{align}
	\setcounter{equation}{6}
	P_{res} & =  {\frac{{P}_{t} {G}_{t}}{{{ 4\pi {{r}^{2}} }}} \times} \frac{S_{ref}}{4 \pi r^2} A_e,
	\end{align}
	where $G_t$ is the antenna gain on the transmitting side, and $A_e$ is the effective area of radar receiver. Applying the equation of receiving gain as 
	
	\begin{align}
	G_{r} & =  {\frac{4 \pi A_e}{\lambda_w^2}}
	\end{align}
	where $L_s$ is the small scale fading of the channels. where $\lambda_w = c/f_w$ is the wavelength of sensing signals with $c$ and $f_w$ being light speed and carrier frequency, respectively. the received power of radar sensor can be transformed as	 
	\begin{align}
	P_{res} & = {P}_{t} {G}_{t} \frac {{S}_{ref}}{{{\left( 4\pi {{r}^{2}} \right)}^{\text{2}}}} \frac{{{G}_{r}}{{\lambda_{\omega} }^{{2}}}}{{4}\pi } \frac{1}{{{L}_{s}}},
	\end{align}
	where the scattering signal loss is $L_s$. }
As analyzed in Section \ref{SubsecID}, the effective JCS detection beam request the SINR for sensing, denoted by $SINR_s$, to be greater than the threshold $\beta_{s}$.
Denoting $\rm Pr_{NLoS} = 1-Pr_{LoS}$ to be the probability of NLoS transmission, the received power of echoes can be derived as
\begin{align}
P_{rs} = P_{res}(n_{p} {{\rm Pr }_{\rm LoS}}+\eta n_{p}{{{\rm Pr }_{\rm NLoS}}{\rm Pr }\{SINR_s\ge\beta_{s}|{\rm NLoS}\}}).
\label{prs}
\end{align}
Due to the attenuation caused by reflection and scattering, the power of the JCS echoes transmitted via the NLoS channel will suffer from severe channel loss. Scattered JCS beams impose additional interference on infrastructure detection\cite{5766057}.
The power of JCS echoes from the NLoS channel is
\begin{align}
{{P}_{rs,\rm NLoS}} 
& = \eta {{n}_{p}}{{P}_{t}}\frac{{{G}_{t}}{{G}_{r}}{{\lambda_{\omega} }^{\text{2}}}{{S}_{ref}}}{{{L}_{s}}4{{\pi }^{\text{3}}}{{r}^{\text{4}}}} (1-{{\rm Pr }_{\rm LoS}}(d,{{h}_{1}},{{h}_{2}})).
\label{PSNLOS}
\end{align}
where the magnitude of $\eta$ is at the order of $10^{-5}$\cite{6043861}. Within urban area, the power spectral density of Gaussian noise is $N = 10^{-11}$\cite{9490660}. The antenna gains are set to $G_t=G_r=8\times16$. The carrier frequency is set to 24 Ghz. The SNR of detection via NLoS channel is less than $2.06$, which is lower than the threshold $\beta_s = 7$ dB for JCS detection. 
In this way, effective JCS detection signals can only be satisfied by echoes from the LoS channel. 
The power of JCS echoes is decisive to successful sensing. The received power of JCS echoes is
\begin{align}
{P}_{rs,\rm LoS} 
= n_p {{\rm Pr }_{\rm LoS}}(d,{{h}_{1}},{{h}_{2}}){{P}_{t}}\frac{{{G}_{t}}{{G}_{r}}{{\lambda_{\omega} }^{\text{2}}}{{S}_{ref}}}{{{L}_{s}}4{{\pi }^{\text{3}}}{{r_s}^{\text{4}}}}.
\end{align}

Then, the expectation of received JCS echo power is
\begin{align}
\label{EPRS1}
E({{P}_{rs,\rm LoS}}) & =E({{n}_{p}}{{P}_{res}}\exp (-{2{r}_{0}}{{\lambda }_{0}}\int_{0}^{d-\frac{\pi }{2}{{r}_{0}}}{ G(h) dx})), 
\end{align}
\begin{figure*}[!t]
	\normalsize
	\begin{align}
	\begin{aligned}
	E({{P}_{rs,LoS}})=&
	d^2 \eta \lambda_0 n_p P_{res} r_0 \sigma_0 \left(3 \pi \lambda_0 r_0 erf \left( \frac{\log \left(-\frac{\pi h {r_0}}{2 d}+h-{\mu_0}\right)}{\sqrt{2} {\sigma_0}}\right)+3 \sqrt{\pi } {\sigma_0}^2 \left(-\sqrt{\pi } {\lambda_0} {r_0}-4 {\mu_0}+2 \sqrt{2 \pi } \sigma_0-2\right) 
	\right.
	\\
	&\left.
	\log ^2\left(-\frac{\pi h {r_0}}{2 d}+h-{\mu_0}\right)+12 {\lambda_0} {r_0} {\sigma_0}^4 e^{-\frac{\log ^2\left(-\frac{\pi h {r_0}}{2 d}+h-{\mu_0}\right)}{2 {\sigma_0}^2}}+4 \sqrt{\pi } {\sigma_0} \left(\sqrt{2} {\lambda_0} {r_0}-2 {\sigma_0}\right) \log ^3\left(-\frac{\pi h {r_0}}{2 d}+h-{\mu_0}\right)
	\right.
	\\
	&\left.
	-3 {\lambda_0} {r_0} \log ^4\left(-\frac{\pi h {r_0}}{2 d}+h-{\mu_0}\right)+6 \pi \sqrt{2} (2 {\mu_0}+1) {\sigma_0}^3 \log \left(-\frac{\pi h {r_0}}{2 d}+h-{\mu_0}\right)/{2 {\sigma_0}^4}\right)/{3 \sqrt{2} \pi }.
	\end{aligned}
	\label{CFS_EPLOS}
	\end{align}
	\hrulefill
	\vspace*{4pt}
\end{figure*}
According to \textbf{Appendix} \textbf{\ref{Appendix:A}}, (\ref{EPRS1}) can be further derived in (\ref{CFS_EPLOS}).

\begin{figure}[t]
	\centering
	\includegraphics[width=1\linewidth]{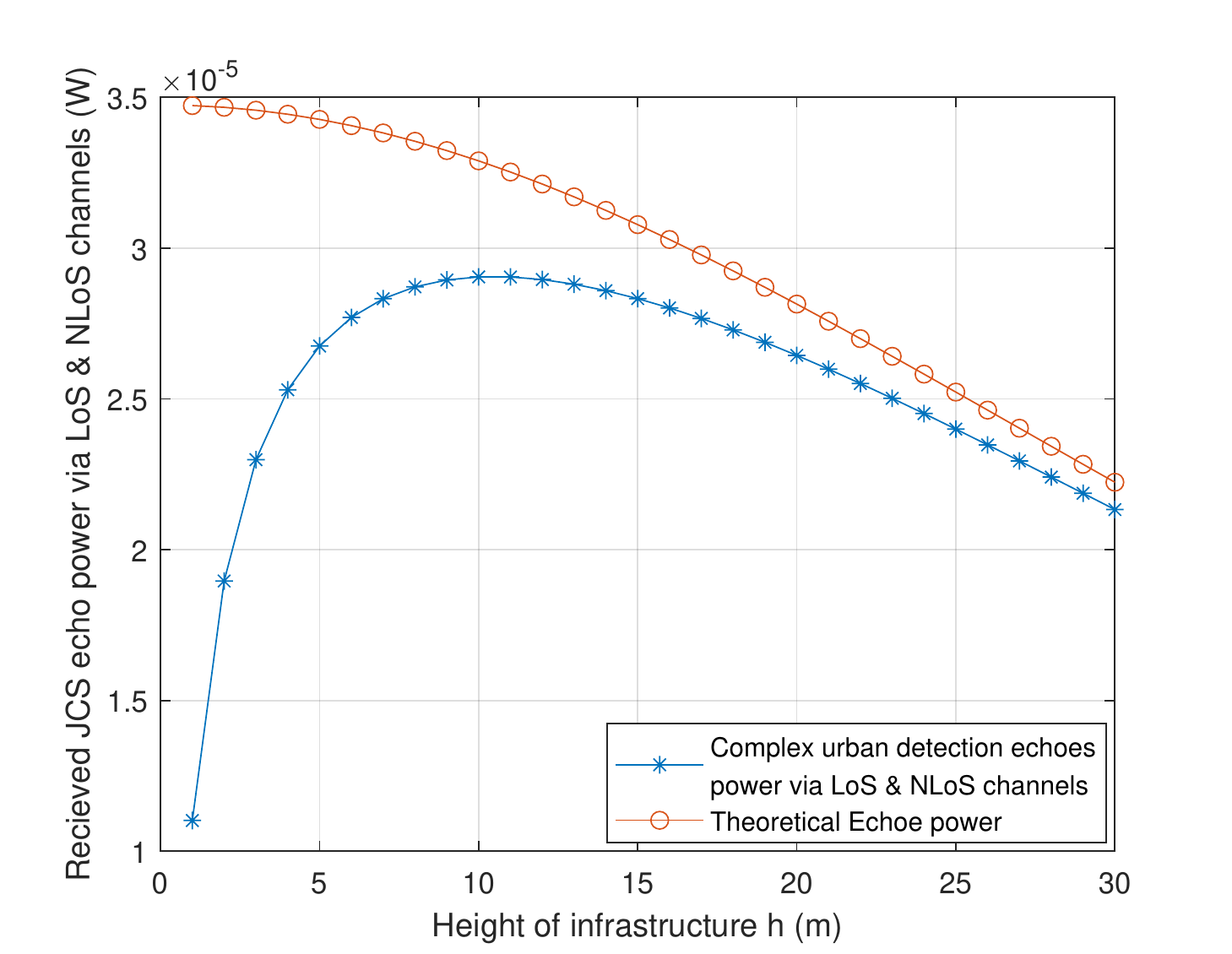}
	\caption{Results of received JCS echoes power.}
	\label{Prsh}
\end{figure}

{Fig. \ref{Prsh} shows the theoretical JCS echoes power from the sensing devices and the simulated JCS echoes power under the complex urban environment with attenuation via LoS and NLoS channels. 
	The simulation data reveals severe attenuation of JCS echoes in urban environments compared with the theoretical results calculated from the radar range equation. Thus, the JCS signal channels in urban areas have different characteristics and the radar range equation is not applicable to correctly modeling the JCS channels.}
\subsubsection{Interference Analysis}
\begin{figure}[t]
	\centering
	\includegraphics[width=1\linewidth]{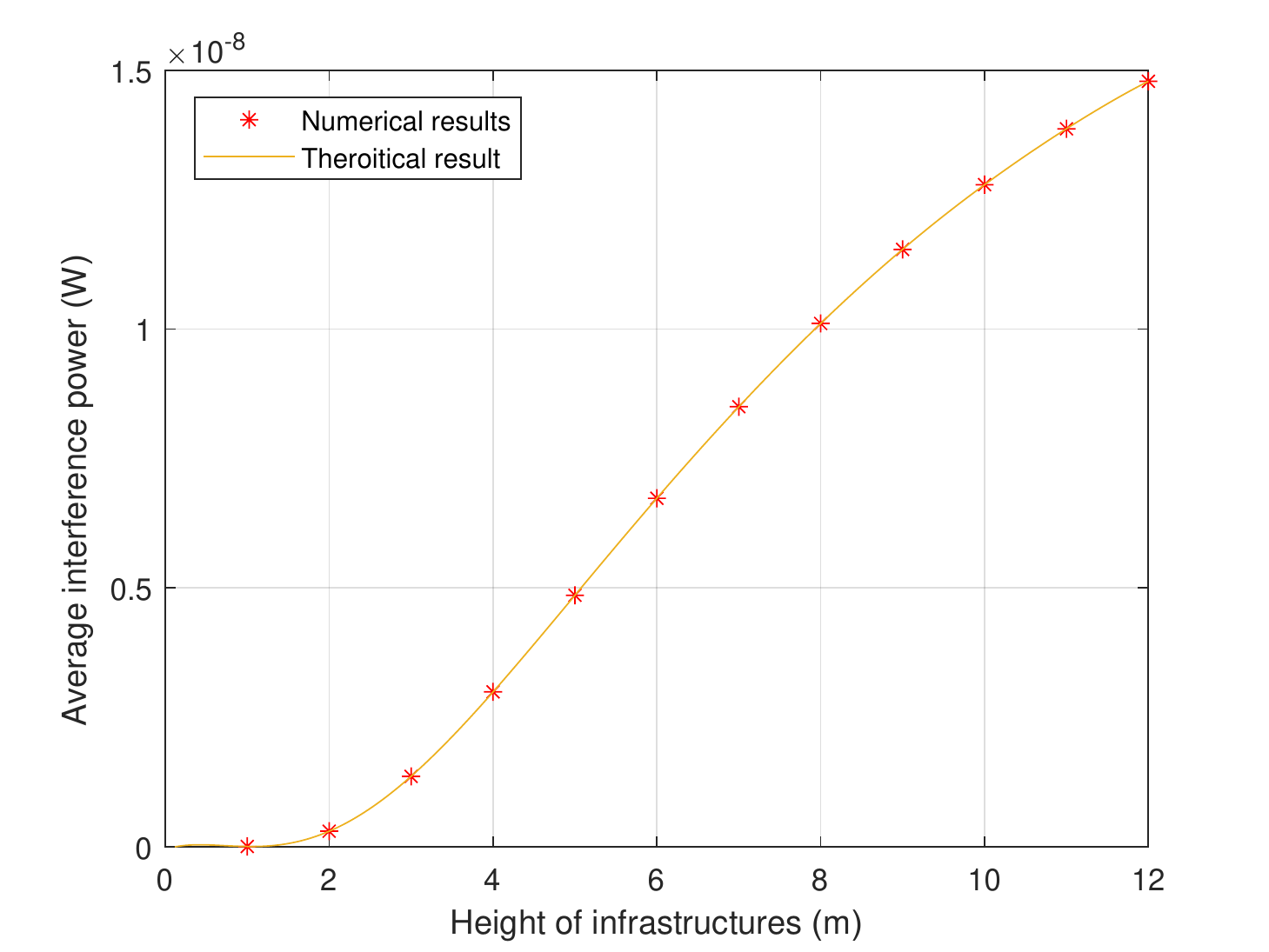}
	\caption{Results of received interference power.}
	\label{Icomm}
\end{figure}

The interference imposed on JCS infrastructures will reduce the SINR of sensing and further deteriorate the JCS detection performance. The interference sources imposed on JCS sensing include Gaussian noise and interference from other wireless devices. The Gaussian noise is assumed to have zero mean and power spectral density, $N=10^{-11}$ W/Hz\cite{9490660}. Besides, interference from other wireless devices includes scattered communication signals from the infrastructures and vehicles, which consist of interfering signals transmitted via LoS and NLoS channels, which can be derived as
\begin{align}
	I_{c}=&{{I}_{c,\rm LoS}}+{{I}_{c,\rm NLoS}} \nonumber\\ 
	=&\sum\limits_{{x_i}\in \Phi_i} {{{P}_{t}}{{\left| x_i \right|}^{-{{\alpha }}}}{{\rm Pr }_{\rm LoS}}(d,{{h}_{1}},{{h}_{2}})} \nonumber\\
	& + \sum\limits_{{x_i \in \Phi_i}} {\eta {{P}_{t}}{{\left| x_i \right|}^{-{{\alpha }}}}(1-{{\rm Pr }_{\rm LoS}}(d,{{h}_{1}},{{h}_{2}}))} , 
\end{align}
where $x_i$ refers to the distance between the $i$-th interference sources and receiver, ${I}_{c,\rm LoS}$ and ${I}_{c,\rm NLoS}$ are the power of interference from remote transmitters via LoS and NLoS channels, respectively. $\Phi_i$ is the position set of interference source transmitters, with distribution following the deployment of JCS road infrastructures.
Fig. \ref{Icomm} illustrates the numerical and theoretical results of interference intensity $I_{c}$. As indicated in Fig. \ref{Icomm}, an increase in gantry height introduces larger interference from remote JCS infrastructures, which will deteriorate the performance of JCS detection and communication. 

In addition, scattered JCS echoes from other JCS infrastructures are also critical interference sources for detecting the typical JCS infrastructure. The power distribution of scattered JCS echoes from the NLoS channel is shown in (\ref{PSNLOS}). In this way, the aggregate power distribution of the interference imposed on JCS detection can be derived as $P_{ns} = I_{c} + P_{rs,\rm NLoS} + N$.
\subsubsection{Probability of Successful Detection}\label{SECPSD}
Successful detection of JCS infrastructures requires $SINR_s$ to be larger than the signal processing threshold. The probability of successful detection is given by
\begin{align}
	 {\rm{ Pr}_{succ, s}}&={\rm Pr}({{SINR}_{s}}\ge {{\beta }_{s}}) ={\rm Pr}(\frac{{{P}_{rs}}}{P_{ns}}\ge {{\beta }_{s}}),
\end{align}
The probability of successful detection can be transformed as\cite{9119440}
\begin{align}\label{Prsuccs}
	{\rm{ Pr}_{succ, s}}
	& = \exp ( - \frac{{\beta_s (N+{P_{rs,\rm NLoS}})}}{{{P_{rs,\rm LoS}}}}) \mathcal{L}_{I_{c}} ( \frac{{\beta_s }}{{{P_{rs,\rm LoS}}}}),
\end{align}
where $\mathcal{L}_{P_f}(\cdot)$ is the Laplace transform of the random signal $P_f$. 
The solution of $\mathcal{L}_{P_f}(\cdot)$ follows \textbf{Theorem} \textbf{\ref{theoLi}}.
\begin{theorem}\label{theoLi}
	For random interference, the Laplace transform of signal power $P_f$ is
	\begin{align}
	{\mathcal{L}_{P_f}}= {E_{{\Phi_i}}}[\prod\limits_{{x_i} \in {\Phi _i}} {\frac{1}{{1 + {{\beta_s }}|{x_i}{|^{ - \alpha }}{\rm{ Pr }_{ch}}}}} ],
	\label{mathcalL1}
	\end{align}
	where ${\rm{ Pr }_{ch}}$ is the probability of random condition for LoS and NLoS channel model. 
	\begin{proof}
		According to the calculation process in \cite{8555855}, the Laplace transform can be derived as
		\begin{align}
		 {\mathcal{L}_{P_f}}&= {E_{{P_f}}}[\exp (\frac{{\beta_s }}{{{P_{t}}}}{P_f})] \nonumber\\
		& = {E_{{\Phi_i}}}[\prod\limits_{{x_i}\in \Phi_i}[\exp g_i(\frac{{\beta_s }}{{{P_{t}}}} {{{P}_{t}}{{\left| x_i \right|}^{-{{\alpha }}}}{{\rm{ Pr }_{ch}}}})]] 	\nonumber\\	 
		&= {E_{{\Phi_i}}}[\prod\limits_{{x_i}\in \Phi_i}E_{g_i}[\exp (\frac{{\beta_s }}{{{P_{t}}}} {{{P}_{t}}{{\left| x_i \right|}^{-{{\alpha }}}}{{\rm{ Pr }_{ch}}}})]] .
		\end{align}
		For base-band signal, the expectation of symbol of JCS detection signal follows $E_{g_i} = 1/N_i \sum_{i=1}^{N_i}g_i=1$. Thus, the solution of ${\mathcal{L}_{P_f}}$ can be derived as (\ref{mathcalL1}).
	\end{proof}
\end{theorem}
Through the calculation in \textbf{Theorem} \textbf{\ref{theoLi}}, the Laplace transform of $I_{c}$ via LoS and NLoS channel, $I_{c,\rm LoS}$ and $I_{c,\rm NLoS}$, can be derived as
\begin{align}\label{LoSNLoSICOMM}
{\mathcal{L}_{I_{c, \rm LoS}}}&= {E_{{\Phi_i}}}[\prod\limits_{{x_i} \in {\Phi _i}} {\frac{1}{{1 + {{\beta_s }}|{x_i}{|^{ - \alpha }}{{\rm Pr }_{\rm LoS}}}}} ], \rm and \nonumber\\
{\mathcal{L}_{I_{c,\rm NLoS}}} 
&= {E_{\Phi_i}}[\prod\limits_{{x_i} \in {\Phi _i}} {\frac{1}{{1 + {{\beta_s }}\eta|{x_i}{|^{ - \alpha }}({{1-\rm \Pr }_{\rm LoS}})}}} ].
\end{align}
\begin{theorem}\label{theoLIcs2}
	As proved, $\mathcal{L}_{I_{c,\rm LoS}}$ and $\mathcal{L}_{I_{c,\rm LoS}}$ can be derived as
	\begin{align}\label{LIcLoS}
	{\mathcal{L}_{I_{c,\rm LoS}}} 
	& = \exp ( - {2\pi\lambda _0} H_{\rm LoS}(\beta_s,d,h)) \nonumber \\
	{\mathcal{L}_{I_{c,\rm NLoS}}} 
	& = \exp ( - {2\pi\lambda _0} H_{\rm NLoS}(\beta_s,d,h)).
	\end{align}
	\begin{proof}
		The probability generating function of $f(x)$ can be derived as \cite{6863654}\cite{7756327}
		\begin{align}\label{probgen}
		E(\prod\limits_{{x_i} \in {\Phi _i}} {f(x))} = \exp ( - {\lambda _0} {[1 - f(x)]dx} ).
		\end{align}
		Applying (\ref{probgen}) to (\ref{LoSNLoSICOMM}), ${\mathcal{L}_{P_f}}$ can be calculated as
		\begin{align}
		{\mathcal{L}_{P_f}} 
		& = \exp ( -2\pi {\lambda _0} \int_{0}^{\infty}(1-{\frac{1}{{1 + {{\beta_s }}\eta|{x_i}{|^{ - \alpha }}({{1- \Pr }_{\rm ch}})}}})).
		\end{align}
		Creating function ${H_{\rm ch}}(\beta_s,d,h)$ as 
		\begin{align}\label{Hfunc}
		H_{\rm ch}(\beta_s,d,h) = \int_0^{\infty} {[1 - {\frac{1}{{1 + {{\beta_s }}|{x_i}{|^{ - \alpha }}{{\rm \Pr }_{\rm ch}}}}}]dx_i},
		\end{align}
		for interference signals from LoS and NLoS channel, ${\mathcal{L}_{I_{c,\rm LoS}}}$ can be derived as (\ref{LIcLoS}).
	\end{proof}
\end{theorem}
Hence, $\mathcal{L}_{I_{c}}$ can be derived as\cite{7412759}
\begin{align}
\mathcal{L}&_{I_{c}}( \frac{{\beta_s }}{{{P_{rs,\rm LoS}}}}) \nonumber \\=&\mathcal{L}_{I_{c,\rm LoS}}( \frac{{\beta_s }}{{{P_{rs,\rm LoS}}}})\mathcal{L}_{I_{c,\rm NLoS}}( \frac{{\beta_s }}{{{P_{rs,\rm LoS}}}}) \nonumber\\
 =& \exp ( - {2\pi\lambda _0} [H_{\rm LoS}(\beta_s,d,h) + H_{\rm NLoS}(\beta_s,d,h)]).
\end{align}
Further calculations about $H_{\rm LoS}$ and $H_{\rm NLoS}$ are shown in \textbf{Appendix} \textbf{\ref{Appendix:B}}.

Based on the analysis of interference from remote infrastructures, the probability of successful detection is
\begin{align}
 &{\rm Pr_{succ,s}}(d)
 = \exp ( - \frac{{\beta_s (N+{P_{rs,\rm NLoS}})}}{{{P_{rs,\rm LoS}}}}) \nonumber\\
&\times \exp ( - {2\pi\lambda _0} [H_{\rm LoS}(\beta_s,d,h) + H_{\rm NLoS}(\beta_s,d,h)]),
\end{align}
\subsubsection{Infrastructure Detection Range}\label{SECIDR}
The performance of infrastructure detection is evaluated by the expectation of detection range, $E({r_{s,\rm max}})$, and is solved in \textbf{Theorem} \textbf{\ref{RDs}}. The expression of $E({r_{s,\rm max}})$ is
\begin{align}\label{EDs}
E&({r_{s,\rm max}}) \nonumber \\ 
=&\sqrt[4]{{\frac{{{n_p}{P_{t}}{G_t}{G_r}{\lambda_w ^{\rm{2}}}{S_{ref}}}}{{{L_s}4{\pi ^{\rm{3}}}{\beta _s}(N + \eta{P}_{res}E(n_p (1-{\rm Pr_{LoS}})) + E({I_{c}}))}}}}.
\end{align}
\begin{theorem}\label{RDs}
	 The maximum detection range is limited by the transmission power, channel loss rate, and noise-interference factor of the detection signal. To meet the requirement of successful detection, the upper bound of the detection range can be derived as
	\begin{align}
	{r_s^{\rm{4}}} \le \frac{{{n_p}{P_{t}}{G_t}{G_r}{\lambda_w ^{\rm{2}}}{S_{ref}}}}{{{L_s}4{\pi ^{\rm{3}}}{\beta _s}(N + {P}_{rs,\rm NLoS} + {I_{c}})}}.
	\end{align}
	\begin{proof}{
		$SINR_s$ can be derived as
		\begin{align}
		SINR_s 
		= {n_p}{P_{t}} \frac{{{G_t}{G_r} {\lambda ^{\rm{2}}} {S_{ref}}}} {{{L_s}4{\pi ^{\rm{3}}}}} \frac{{E(n_p\rm Pr_{LoS})}} {{N + {I_{c}}+P_{rs,\rm NLoS}}}.
		\end{align}
		As presented in Section \ref{SECII}, ${P}_{rs,\rm NLoS}$ and ${I_{c}}$ are independently and identically distributed. The set of coverage area is denoted by $R_s\{i,j\}=\{\Phi_i|SINR_s\{i,j\}\ge\beta_s, \{i,j\} \in \Phi_i\}$, where $\Phi_i$ is the set of area under coverage, $i$ and $j$ are area coordinates. The expectation of $I_{c}$ can be derived as
		\begin{align}
		E(I_{c})=&E(\sum\limits_{{x_i}\in \Phi_i} {{{P}_{t}}{{\left| x_i \right|}^{-{{\alpha }}}}{\rm {Pr }_{LoS}}(d,{{h}_{1}},{{h}_{2}})}) \nonumber\\
		& + E(\sum\limits_{{x_i \in \Phi_i}} {\eta {{P}_{t}}{{\left| x_i \right|}^{-{{\alpha }}}}(1-{\rm {Pr }_{LoS}}(d,{{h}_{1}},{{h}_{2}}))}).
		\end{align}
		Applying Campbell's theorem for sums, where $E(S) = \int_{R^d}f(x)\lambda(x)dx$ when $S = \sum\limits_{x\in\Phi}f(x)$, for expectation of power $P_f$ as $E_{P_f}$,
		\begin{align}
		E_{P_f}
		=& {\lambda _d}\int_0^{2\pi } {\int_{{R_{coop}}}^{ + \infty } {{P_t}{{\left| {{x_i}} \right|}^{ - \alpha }}{\rm Pr_{ch}}(d,{h_1},{h_2})} } dx,
		\end{align}
		where $\lambda_d$ is the density of devices and $R_{coop}$ is the cooperative detection range. Thus, the expectation of $I_{c}$ is
		\begin{align}
		E&(I_{c}) \nonumber \\
		=& {\lambda _d}\int_0^{2\pi } {\int_{{R_{coop}}}^{ + \infty } {\eta{P_t}{{\left| {{x_i}} \right|}^{ - \alpha }} (1-{\rm Pr_{LoS}}(d,{h_1},{h_2}))} } dxd\theta \nonumber\\
		& +{\lambda _d}\int_0^{2\pi } {\int_{{R_{coop}}}^{ + \infty } {{P_t}{{\left| {{x_i}} \right|}^{ - \alpha }}{\rm Pr_{LoS}}(d,{h_1},{h_2})} } dxd\theta.
		\end{align}}
	\end{proof}
\end{theorem}
\subsection{Performance of I2V Communication}\label{COMM}
\subsubsection{Attenuation and interference analysis of I2V communication}
The received power of JCS I2V communication consists of additive power from beams via both LoS and NLoS communication channels. The received power is denoted by $P_{rc}$, and is given by
\begin{align}
{{P}_{rc}} =&{{P}_{rc,\rm LoS}}+{{P}_{rc,\rm NLoS}} \nonumber\\ 
=&{{P}_{t}}G_t{{\left| x \right|}^{-{{\alpha }}}}g_i G_{rv}{\rm {Pr }_{LoS}}(d,{{h}_{1}},{{h}_{2}})\nonumber\\
& +\eta {{P}_{t}}G_t{{\left| x \right|}^{-{{\alpha }}}}g_iG_{rv}(1-{\rm {Pr }_{LoS}}(d,{{h}_{1}},{{h}_{2}})).
\end{align}
Interference is the other critical factor that influences the performance of JCS I2C communication. Gaussian noise $N$ and inter-device interference $I_{c}$ are still included in the interference imposed on I2V communication. Besides, both scattered and reflected echo beams from LoS and NLoS channels will impose interference on JCS communication, the power of which is denoted by $I_{ref}$. As provided in Section \ref{detectionsignal}, $I_{ref}$ includes beam power of LoS and NLoS channels, which is expressed by combining (9) and (10) as
\begin{align}
I_{ref} &=P_{rs,\rm LoS} + P_{rs,\rm NLoS}\nonumber\\
&= n_p {{P}_{t}}\frac{{{G}_{t}}{{G}_{r}}{{\lambda_{\omega} }^{\text{2}}}{{S}_{ref}}}{{{L}_{s}}4{{\pi }^{\text{3}}}{{r_s}^{\text{4}}}}(\rm Pr_{\rm LoS}+\eta (1-\rm Pr_{\rm LoS})).
\end{align}
At last, the aggregate interference is given by
\begin{align}
	P_{nc} = N+I_{c}+I_{ref}.
\end{align}
\subsubsection{Probability of successful I2V communication}\label{SECIAC}
Successful I2V communication requires the received signal to be larger than the SINR threshold $\beta_c$, i.e. as $SINR_c\ge\beta_c$. Then, The probability of successful I2V communication can be expressed as
\begin{align}
&{\rm {Pr}_{succ,c}} = {\rm Pr}(\frac{{{P}_{rc}}}{N+{{I}_{{c}}}+{{I}_{ref}}}\ge \beta_c ) \nonumber\\ 
=&\exp (-\frac{\beta_c (N)}{{{P}_{rc}}}){{\mathcal{L}}_{{{I}_{\text{comm}}}}}(\frac{\beta_c }{{{P}_{rc}}}){\exp }(-\frac{\beta_c ({{I}_{ref}})}{{{P}_{rc}}}),
\end{align}
where $\mathcal{L}_{I_{c}}$ is
\begin{align}
\mathcal{L}_{I_{c}}(\frac{\beta_c}{{{P}_{rc}}})
=& \exp ( - {2\pi\lambda _0} [H_{\rm LoS}(\beta_c,d,h_1,h_2)\nonumber\\& + H_{\rm NLoS}(\beta_c,d,h_1,h_2)]).
\end{align}
Hence, the probability of successful communication can be expressed as
\begin{align}\label{PRSSUCCC}
&{\rm {Pr}_{succ,c}}(d)=\exp (-\frac{\beta_c (N)}{{{P}_{rc}}}) {\exp }(-\frac{\beta_c ({{I}_{ref}})}{{{P}_{rc}}}) \nonumber\\
&\times\exp ( - {2\pi\lambda _0} [H_{\rm LoS}(\beta_c,d,h_1,h_2) + H_{\rm NLoS}(\beta_c,d,h_1,h_2)]),
\end{align}
\subsubsection{JCS communication range}
The communication range for JCS road infrastructures is defined as the maximum range that makes $Pr_{succ,s}$ larger than a threshold of $80\%$. $SINR_c$ can be derived as
\begin{align}
S&INR_c\nonumber\\
 =& \frac{{{{P}_{t}}{{\left| x \right|}^{-{{\alpha }}}}g_i({\eta (1-{\rm {Pr }_{LoS}}(d,{{h}_{1}},{{h}_{2}}))}+
{\rm {Pr}_{LoS}}(d,{{h}_{1}},{{h}_{2}}))}} {{N + {I_{c}}+I_{ref}}}.
\end{align}
Hence, the range of communication is
\begin{align}
 {{\left| r_c \right|}^{{{\alpha }}}}\le\frac{{{{P}_{t}}g_i({\eta (1-{\rm {Pr }_{LoS}}(d,{{h}_{1}},{{h}_{2}}))} + {\rm {Pr}_{LoS}} (d,{{h}_{1}},{{h}_{2}}))}} {\beta_c({N + {I_{c}}+I_{ref}})}.
\end{align}
The expectation of $I_{c}$was shown in Section \ref{SECIAC}, and the expectation of $I_{ref}$ is
\begin{align}
E({I_{ref}}) &= E({P_{rs,\rm LoS} + P_{rs,\rm NLoS}})\nonumber\\
&= \eta{P}_{res}E(n_p (1-{\rm Pr_{LoS}})) + \eta{P}_{res}E(n_p {\rm Pr_{LoS}}).
\end{align}
In this way, the expectation of communication range is
\begin{align}
E(r_{c,\rm max}^{{{\alpha }}})=\frac{{{{P}_{t}}g_i({\eta} + (1-\eta) {\rm {Pr}_{LoS}} (d,{{h}_{1}},{{h}_{2}}))}} {\beta_c({N + E({I_{c}})+E(I_{ref})})}.
\end{align}
\subsection{Performance of I2V Cooperative Detection}\label{ANAPRCOOP}
The performance of cooperative detection depends on both JCS detection and I2V communication. The probability of successful JCS detection and information sharing follows the joint probability of successful JCS detection on target while successful JCS tracking and communication on vehicles. Denoting the case of successful detection on targets, vehicles and I2V communication as $\phi_{s,t}$, $\phi_{s,v}$, $\phi_{c,v}$, respectively. Further denoting $x_{t}=\sqrt{(h)^2+(d_{T})^2}$ and $x_v=\sqrt{(h)^2+(d_{v})^2}$ are the straight line distance between infrastructures and target or vehicles respectively. In this way, the probability of successful cooperative detection is
\begin{align}
	{\rm {Pr}_{coop}}&={\rm Pr}\{\phi_{c,v}\phi_{s,t}\phi_{s,v}\}.
\end{align}
Using different beams for JCS detection, $\phi_{s,t}$ is independent with $\phi_{s,v}$ and $\phi_{c,v}$, so there is
\begin{align}
	{\rm {Pr}_{coop}}={\rm Pr}\{\phi_{s,t}\}Pr\{\phi_{c,v}|\phi_{s,v}\}Pr\{\phi_{s,v}\}.
\end{align}
According to (\ref{Prsuccs}) and (\ref{PRSSUCCC}), ${\rm Pr}\{\phi_{s,t}\}={\rm{ Pr}_{succ,s}}({x_{T}})$, ${\rm Pr}\{\phi_{s,v}\}={\rm{ Pr}_{succ,s}}({x_{v}})$, ${\rm Pr}\{\phi_{c,v}\}={\rm{ Pr}_{succ,c}}({x_{v}})$, and ${\rm Pr}\{\phi_{c,v}|\phi_{s,v}\}={\rm Pr}\{SINR_c\ge\beta_c|SINR_s\ge\beta_s\}$. In this way, 
\begin{align}
{\rm Pr_{coop}}=&{\rm Pr}\{SINR_c\ge\beta_c|SINR_s\ge\beta_s\}\nonumber\\&\times {\rm{Pr}_{succ,s}}(x_{v}){\rm {Pr}_{succ,s}}(x_{T}).
\end{align}
Setting $\tau=\beta_s/\beta_c$, the probability of ${\rm Pr}\{SINR_c\ge\beta_c|SINR_s\ge\beta_s\}{\rm{Pr}_{succ,s}}(x_{v})$ equals to
\begin{align}
	&{\rm Pr}\{SINR_c\ge\beta_c|SINR_s\ge\beta_s\}{\rm{Pr}_{succ,s}}(x_{v})\nonumber\\
	&={\rm Pr}\{SINR_c\ge \tau SINR_s\} {\rm {Pr}_{succ,s}}(x_v)\nonumber\\
	&+{\rm Pr}\{\tau SINR_s\ge SINR_c\} {\rm {Pr}_{succ,c}}(x_v).
\end{align}
To obtain the numerical result of the JCS cooperative detection performance, an algorithm is proposed by solving the range of JCS cooperative detection and coverage probability as \textbf{Algorithm \ref{alg:solvecooprange}}.
Setting $\gamma_{co}$ as the threshold of successful cooperation, I2V cooperation range $d_v$ is calculated at controlled infrastructure height $h$. Besides, by variable controlling, the effective target detection range under cooperation $d_T$ is also calculated. {After that, \textbf{Algorithm 1} uses 2 process to obtain the numerical results, the probability of successful cooperative detection with the correlated $h$ and $d_v$ for each typical target distance $d_T$, and JCS cooperation coverage probability on the tested road sectors.} 
\begin{algorithm}[htb] 
	\caption{Numerical calculation of JCS cooperative detection range and coverage probability.}
	\begin{algorithmic}[1] 
		\Require 
			Number of vehicle and targets, $N_v$ and $N_T$.
			Distance to target, $d_T$,
			to vehicles, $d_v$,
			Height, $h$.
		\Ensure 
			\For{i = 1:$N_v$, j = 1:$N_T$}
				\State 
				$(h,d_v)$ = solve \{min($SINR_s$\{$h$,$d_v$\}, $SINR_c$ \{$h$,$d_v$\}/ ($\beta_c$/$\beta_s$), $SINR_s$ \{ $h$,$d_T$ \}) $\ge$ $\beta_s$ \};
			\EndFor	
		\For{i = 1:$N_v$}
		\State 
		${\rm{Pr}_{cover}}(h,m_{v},\nu_{v})$=solve(${\rm Pr}\{(\mathcal{R}) in (d_v)\}\ge\gamma_{co}$);
		\EndFor	\\
		\Return $h$, $d_v$, ${\rm{Pr}_{cover}}$
	\end{algorithmic} 
\label{alg:solvecooprange}
\end{algorithm} 

{To analyze the complexity of \textbf{Algorithm 1}, We denote the $n_{data}$ to be the number of  datasets of the datapoint parameters. The datapoints store the measured data of vehicles and obstacles. 
	Each datapoint have 3 parameters, height $h$, radius $r_0$ and the distance from the point to the infrastructures, $d_T$ and $d_v$, for vehicle and obstacle, respectively. There are two progresses in \textbf{Algorithm 1}, the first process calculates the probability of successful cooperative detection, and the second process count the datapoints that successfully cooperates with the infrastructure and calculates the coverage probability numerically. The complexity of \textbf{Algorithm 1} is defined as the sum of complexity of the two processes.
	
	The complexity of the first process is as follows:
	
	1. The $SINR_s\{h,d_v\}$, $SINR_c\{h,d_v\}$ and $SINR_s\{h,d_T\}$ are calculated with complexity $\mathcal{O}\{n\}$
	
	2. The minimum of $SINR_s\{h,d_v\}$, $SINR_c\{h,d_v\}$ and $SINR_s\{h,d_T\}$ with each $d_T$ and $d_v$ are selected with complexity $\mathcal{O}\{3n_{data}\}$
	
	3. Using the result of step 2, the number of points with minimum SINR larger than threshold are counted with complexity $\mathcal{O}\{n_{data}\}$
	
	In this way, the complexity of the first process is $\mathcal{O}\{n_{data}+n_{data}+3n_{data}\}$
	
	After completing the first process, the result is used to calculate the coverage probability of JCS cooperative detection. The second process traverse each datapoint in the set of distance data $\mathcal{R}$ and obtain the the probability of successful cooperative detection at each point. The simulation runs $n_{data}$ round, and the complexity of the second process is $\mathcal{O}\{n_{data}^2$\}
	
	Thus, the complexity of \textbf{Algorithm 1} is $\mathcal{O}\{n_{data}^2+5n_{data}\} = \mathcal{O}\{n_{data}^2\}$}

\section{Simulation Results and Analysis}\label{SECIV}
In this section, the performance of road infrastructure cooperative detection is studied. The result of probability analysis on JCS LoS transmission, successful JCS detection and I2V communication is proposed in Section \ref{SimPrLoS}, \ref{SimPrsuccs} and \ref{SimPrsuccc}, respectively. The SINR of JCS detection and I2V communication under different infrastructure deployment schemes are compared based on the distribution of vehicles and obstacles proposed in Section \ref{ObsANA}. Then, the probability of successful JCS detection and I2V communication is analyzed based on the infrastructure deployment scheme. {An infrastructure height that meets the JCS detection and I2V communication range requirements is selected, and the probability of successful JCS detection and I2V communication is then analyzed with different obstacle densities.} The probability of successful cooperative detection is analyzed with different distances of JCS detection and I2V communication. Finally, the numerical result of coverage probability under different infrastructure deploy heights is proposed.

{The simulation parameters are set as follows. Based on the newest vehicular antenna technology from National Instruments\cite{VATNA}, we set the size of uniform plane antenna arrays equipped on connected vehicles as $2\times 16$. Besides, based on the antenna array design from \cite{8246333}, the $8\times 16$ antenna array is a potential solution for future road infrastructure antenna design, so in this paper, we set the sizes of antenna arrays on road infrastructures for JCS cooperative detection as $8\times 16$. Other JCS signal parameters are set according to the empirical data and solution from \cite{9359665} where the base-band power is set as $1$ W, and the SINR thresholds for the signal processing devices of JCS detection and communication is $\beta_{s}=7$ dB and $\beta_c = 13$ dB respectively. The carrier frequency of infrastructures is $24$ GHz referenced to \cite{3GPPTR37}. The normalized RCS of targets is applied as $S_{ref}=1$ $\rm m^2$ according to \cite{8246333}.}
\subsection{LoS Transmission Probability}\label{SimPrLoS}
\begin{figure}[t]
	\centering
	\includegraphics[width=1\linewidth]{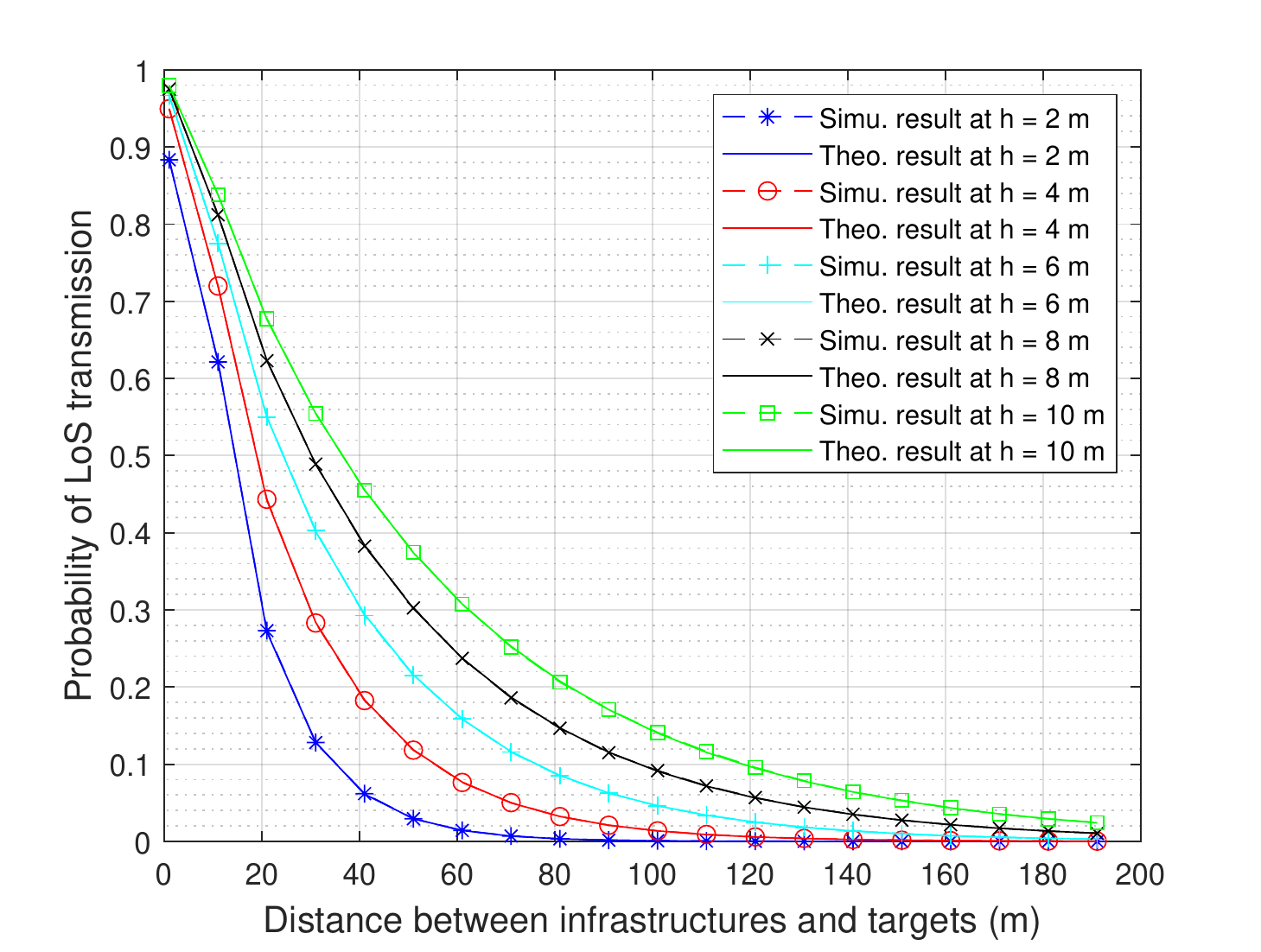}
	\caption{Probability of LoS transmission at different infrastructure height.}
	\label{PrLoS}
\end{figure}

The LoS transmission probability for JCS signals is shown in Fig. \ref{PrLoS}. An increase in transmission distance imposes more obstruction to JCS signals. Thus $\rm Pr_{LoS}$ declines with the increase of the transmission distance of JCS signals. 
Moreover, the increase in infrastructure height mitigates obstructions, and $\rm Pr_{LoS}$ increases. Therefore, proper height adjustment of the antenna array significantly reduces the performance fluctuation caused by obstruction to JCS signals and further improves the JCS cooperative detection performance.
\subsection{Successful Detection Probability}\label{SimPrsuccs}
\begin{figure}[t]
	\centering
	\includegraphics[width=1\linewidth]{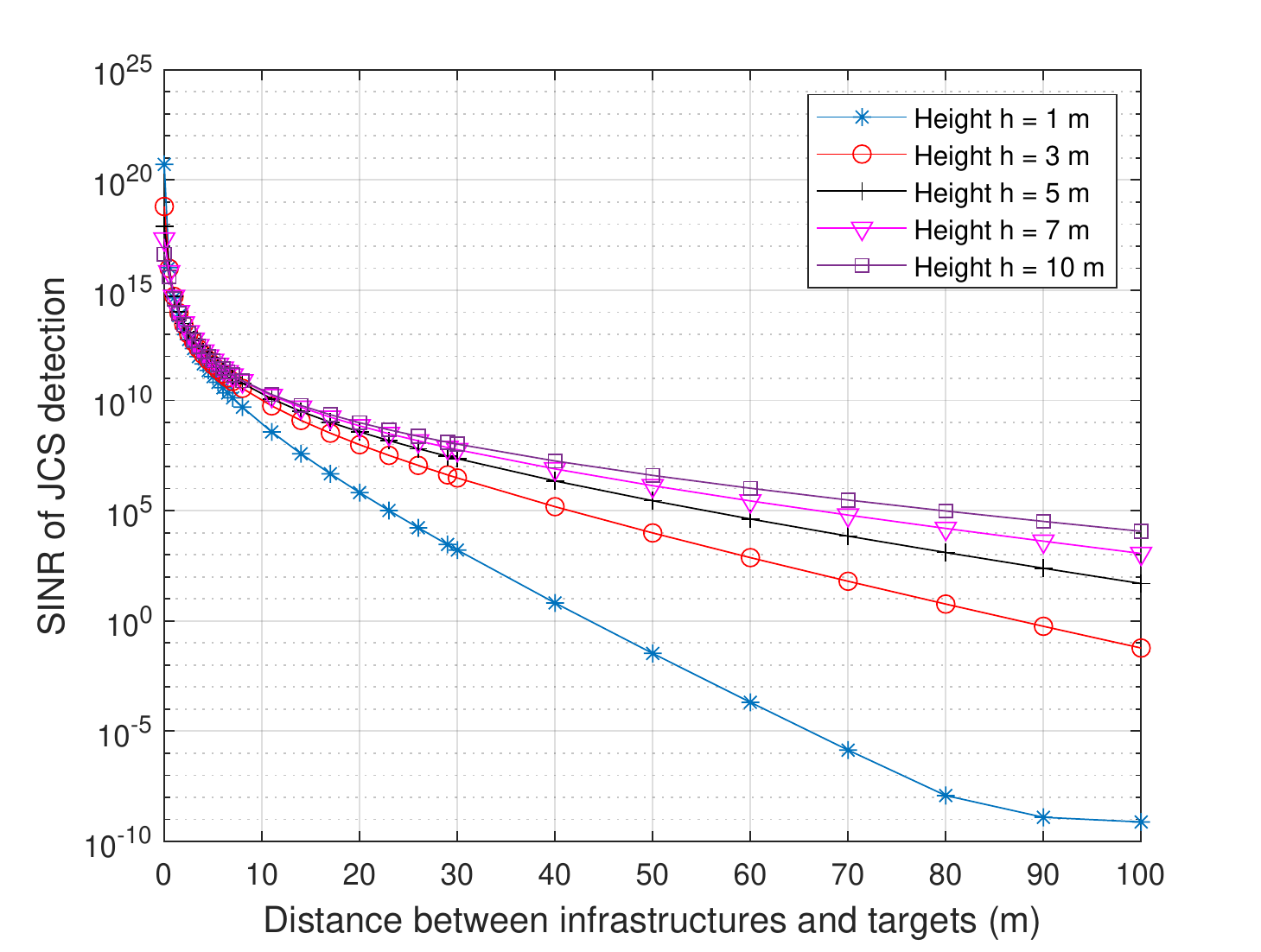}
	\caption{SINR of JCS detection at different infrastructure height.}
	\label{SINRs2}
\end{figure}

Fig. \ref{SINRs2} shows the SINR of JCS detection, denoted by $SINR_s$, which also decrease with the distance between the infrastructures and the targets.
 As shown in Fig. \ref{SINRs2}, the fading of JCS echoes arises due to obstruction and severely reduces $SINR_s$. When the deployment height of the antenna array is 1 m, $SINR_s$ will drop below the threshold of $\beta_{s} = 7$ dB with a detection distance of less than 40 m. Adjusting the infrastructure height will reduce the probability of signal obstructions and mitigate the fading of JCS echoes. $SINR_s$ will be kept at the level of $10^3$ when the antenna height is 10 m and the target's distance is 100 m. 

\begin{figure}[t]
	\centering
	\includegraphics[width=1\linewidth]{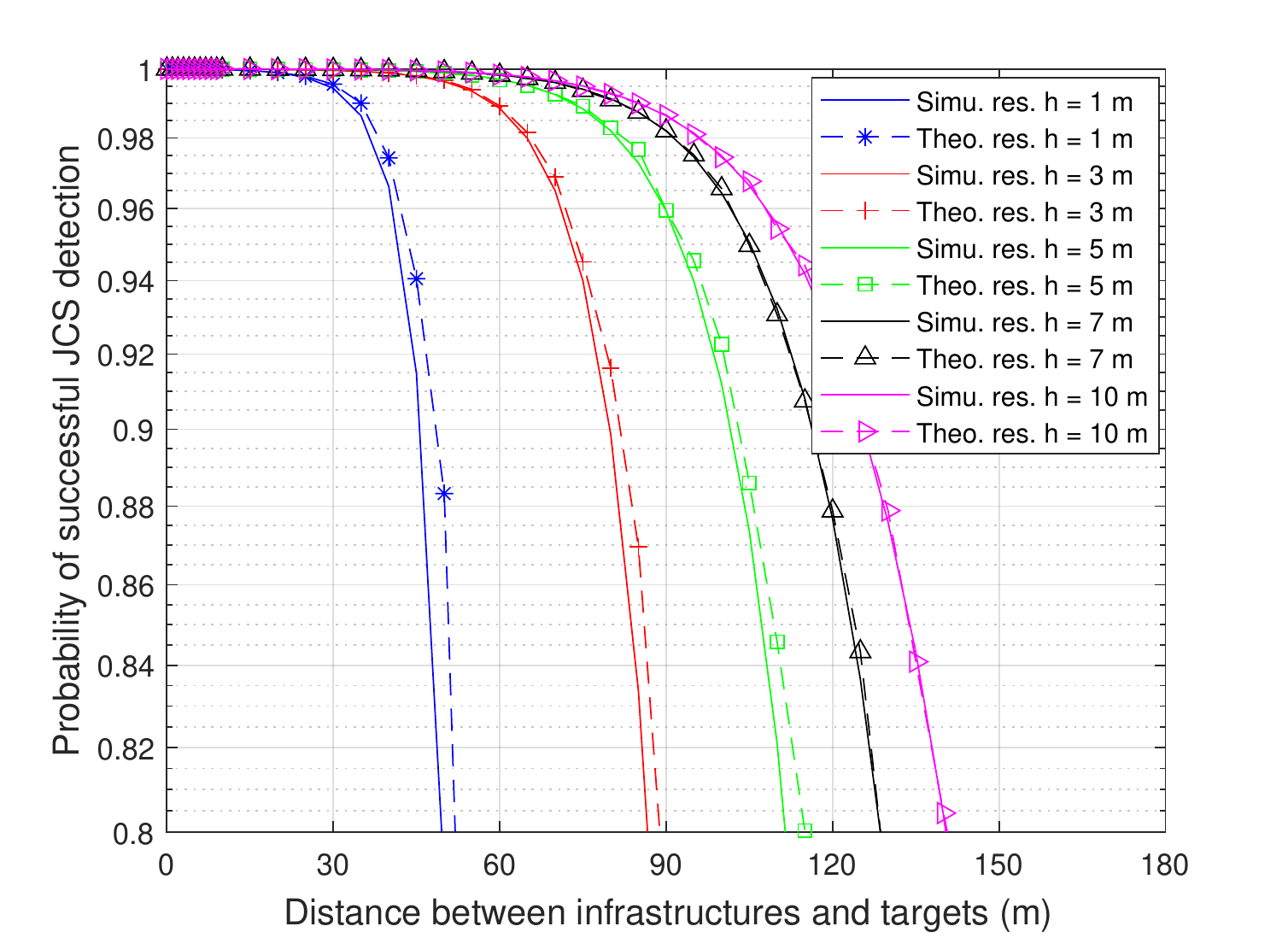}
	\caption{Probability of successful detection at different infrastructure height.}
	\label{PrsuccessS}
\end{figure}

Fig. \ref{PrsuccessS} plots the probability of successful JCS detection, $\rm Pr_{succ,s}$, which changes with the distance between infrastructures and targets. 
The decline rate of $\rm Pr_{succ,s}$ becomes larger with the increase in the distance between infrastructures and targets.{ When $\rm Pr_{succ,s}$ decrease to 80\%, the reliability of detection will not meet the requirement of JCS road infrastructures. In this way, the results of detection distance with a successful detection probability lower than 80\% are omitted to better explain the result that meets the detection requirements. The analysis of communication and cooperative detection uses the same method as this part of the explanation.} {As shown in Fig. \ref{PrsuccessS}, infrastructures deployed at higher position reduces the channel fading caused by signal obstruction and result in a larger range of successful detection. However, when the deploy height ranges from 7 m to 10 m, the JCS detection range only increases by less than 15\%. The effect of detection range extension by further increasing the infrastructure height is limited, and the overtop deployment may cause other problems to JCS cooperative detection, such as a reduction in coverage probability.}

\begin{figure}[t]
	\centering
	\includegraphics[width=1\linewidth]{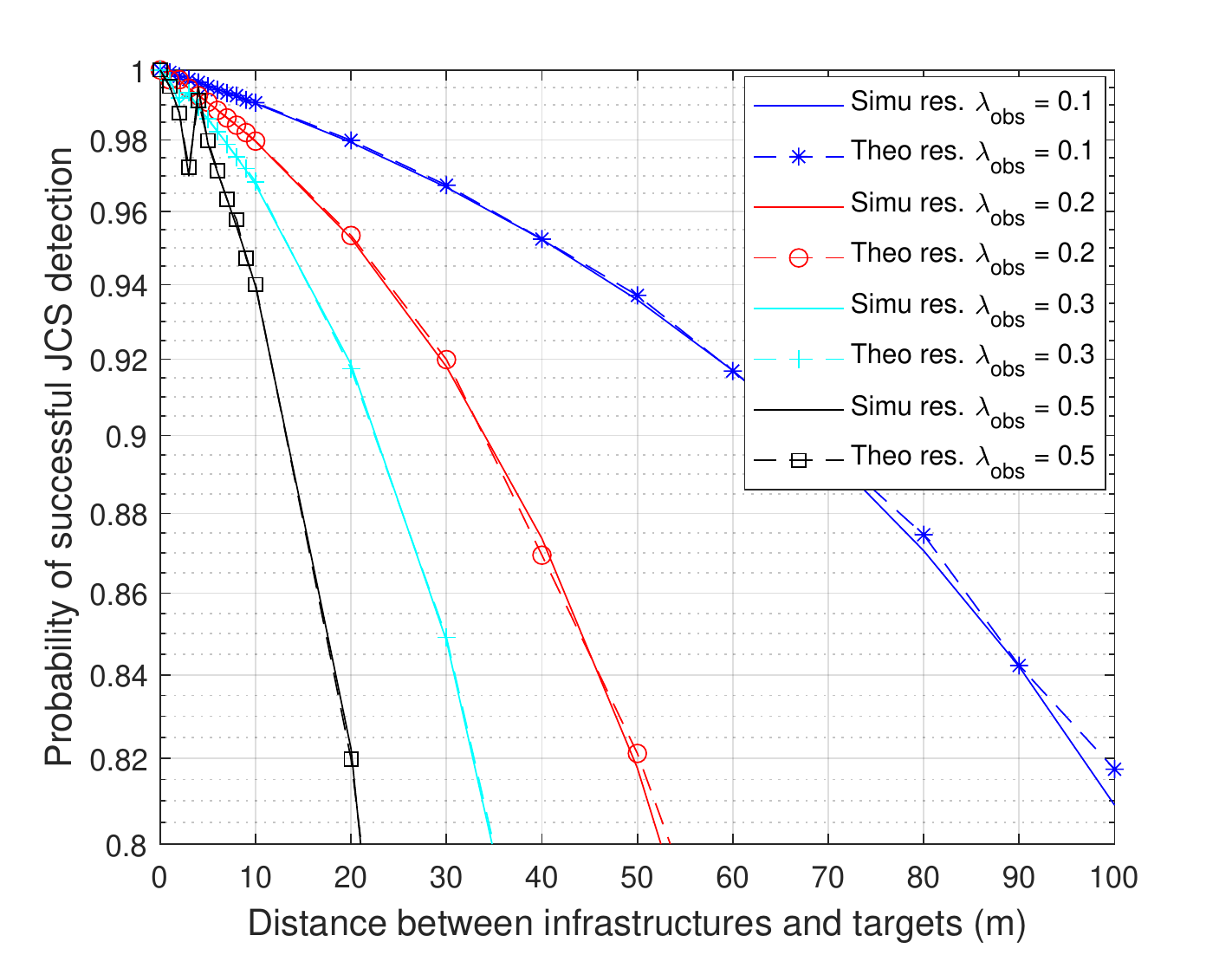}
	\caption{Probability of successful detection with different obstacle density.}
	\label{PrsuccessSobs}
\end{figure}

{Beyond that, we simulated the probability of successful JCS detection, which is shown in Fig 12, where $\rm Pr_{succ,s}$ changes with the distance between infrastructures and targets under different obstacle densities. We selected the obstacle position sets with the density $\lambda_{obs}$ of obstacles ranging from 0.1 to 0.5, where $\lambda_{obs} = 0.1$ represents the slow vehicle flow situation, $\lambda_{obs} = 0.2$ is the jammed or intersection waiting model, and larger $\lambda_{obs}$ represents the complex urban vehicles and obstacles distribution that causing heavy signal obstructions. The infrastructure height is set to 6 m. As shown in Fig. \ref{PrsuccessSobs}, the increase of obstacle density $\lambda_{obs}$ severely reduces the probability of successful JCS detection, which is caused by signal obstructions and attenuation.}

\subsection{Successful Communication Probability}\label{SimPrsuccc}
\begin{figure}[t]
	\centering
	\includegraphics[width=1\linewidth]{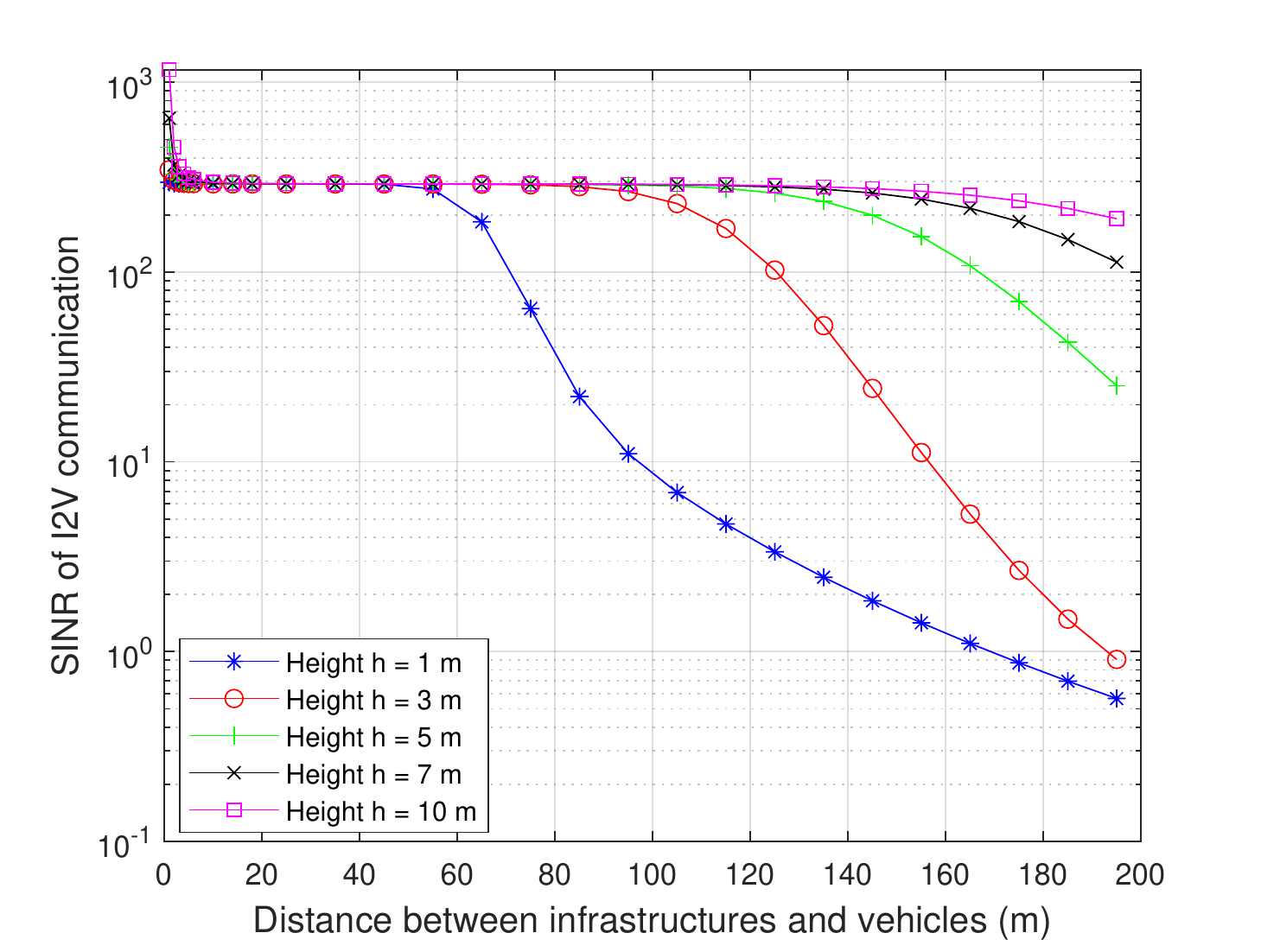}
	\caption{SINR of I2V communication at different infrastructure height.}
	\label{SINRc}
\end{figure}
{Fig. \ref{SINRc} gives the SINR of I2V communication, $SINR_c$, changing with the distance between infrastructures and vehicles, based on the distribution of vehicles from empirical statistics presented in Section \ref{ObsANA}. 
	As shown in Fig. \ref{SINRc}, the $SINR_c$ maintains a high level with the I2V distance between 0 m to 60 m. After that, $SINR_c$ begin to decrease at different I2V distances. Lower deployed infrastructure suffers $SINR_c$ decrease at closer communication range. Increasing the height of infrastructures obviously delays the decrease of $SINR_c$, which ensures a larger communication range and higher successful communication probability.}
\begin{figure}[t]
	\centering
	\includegraphics[width=1\linewidth]{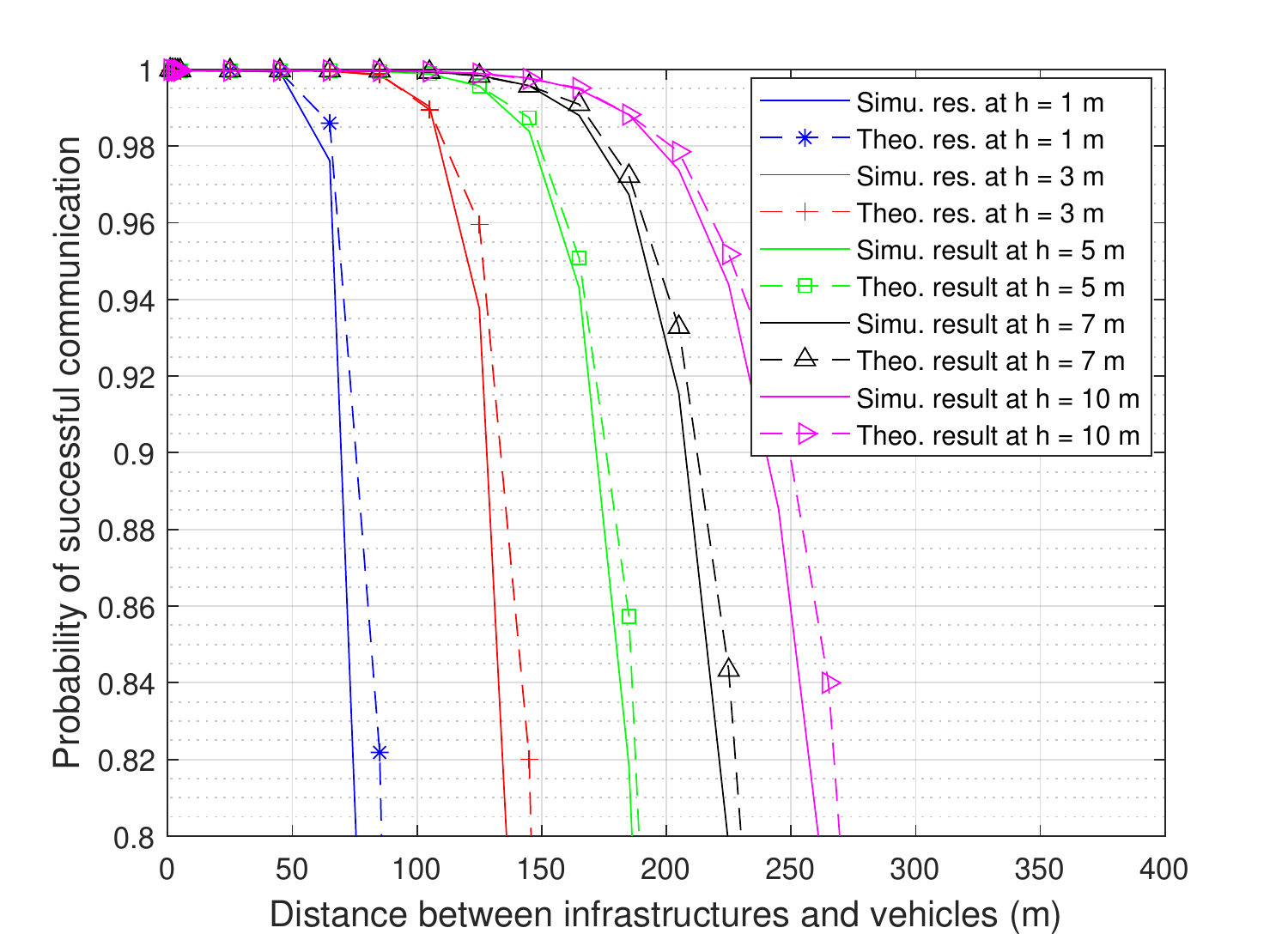}
	\caption{Probability of successful communication at different infrastructure height.}
	\label{PrsuccessC}
\end{figure}

{Fig. \ref{PrsuccessC} presents the probability of successful I2V communication, $\rm Pr_{succ,c}$, which changes with the distance between infrastructures and vehicles. The probability of successful I2V communication is obtained by simulating multiple I2V links of the infrastructure. As shown in Fig. \ref{PrsuccessC}, same as the trend of successful JCS detection, the probability of successful I2V communication decreases when the I2V distances increase. Besides, increasing the deploy height of JCS infrastructures also enlarges the probability of successful communication. However, different from JCS detection, the effect of communication range increases declines much slower than the probability of successful JCS detection. It can be concluded that I2V communication suffers less attenuation with the non-sparse channel characteristics and the ability of multi-path transmission. However, the asynchronicity of the increase rate for JCS detection and communication range may impose further problems of coverage probability, which means the height adjustment of JCS infrastructures has a limited range.}

\begin{figure}[t]
	\centering
	\includegraphics[width=1\linewidth]{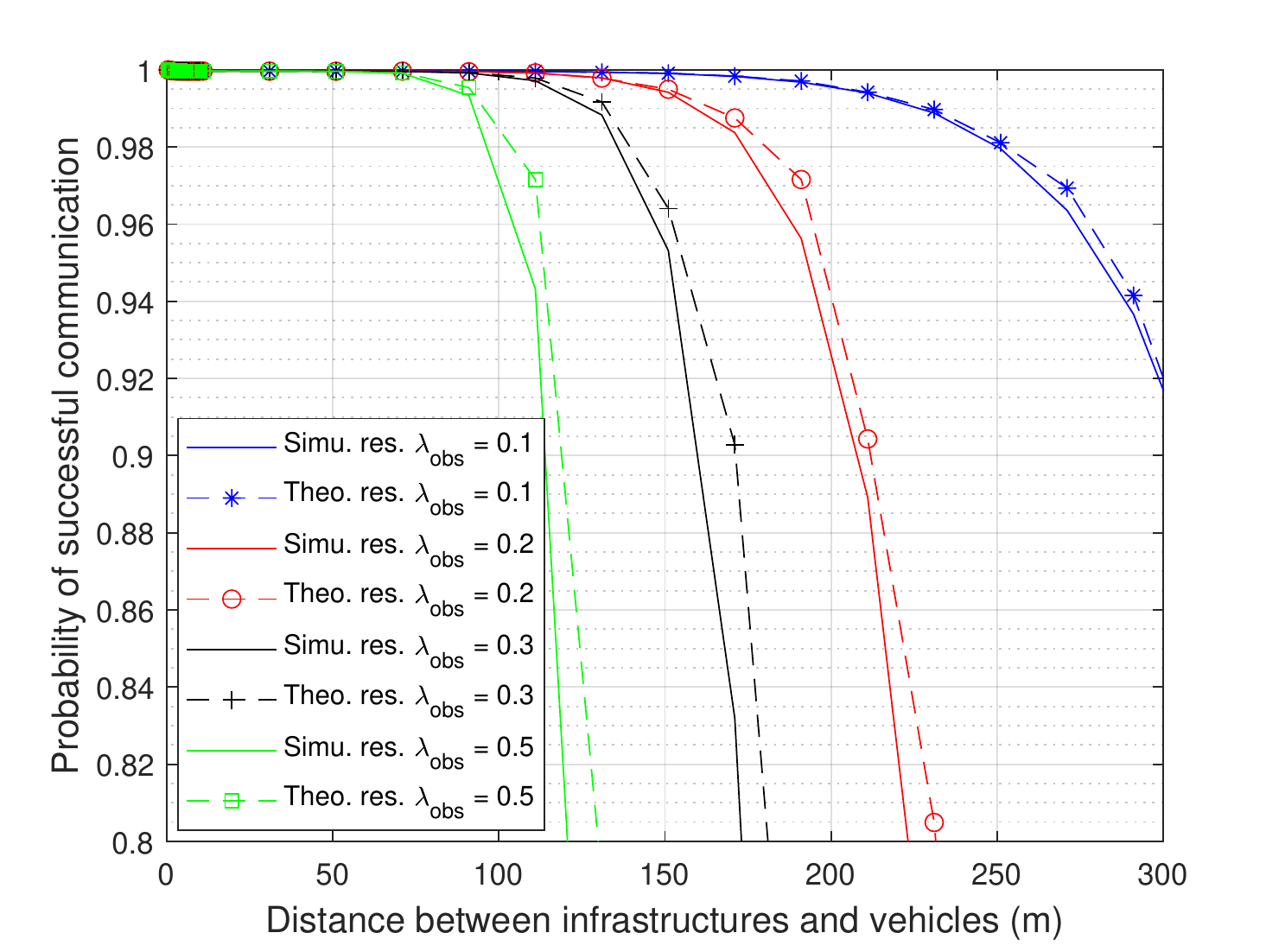}
	\caption{Probability of successful communication with different obstacle density.}
	\label{PrsuccessCobs}
\end{figure}
{Fig. \ref{PrsuccessCobs} shows $\rm Pr_{succ,c}$ with variant obstacles density$\lambda_{obs}$. The $\lambda_{obs}$ was selected at the same value as the simulation for JCS detection to keep the uniformity of the results. As shown in Fig. \ref{PrsuccessCobs}, the $\rm Pr_{succ,c}$ also decreases with larger $\lambda_{obs}$ employed. However, compared with the changing trend of $\rm Pr_{succ,s}$, the ratio of $\rm Pr_{succ,c}$ decrease is also smaller, which means communication has better robustness than JCS detection signals. However, when $\lambda_{obs} = 0.5$, which means the most complex urban obstruction situation, the range of I2V communication is similar to the range of JCS detection. In this way, it could be proved that under complex obstructed LoS \& NLoS channels situation, the range of both JCS detection and communication is limited to the obstacle distance due to channel sparsity.}
\subsection{Cooperative Detection Performance}\label{SimPrcoop}
As analyzed within Section \ref{ANAPRCOOP}, the probability of successful cooperative detection ($\rm Pr_{coop}$) is influenced by I2O, I2V detection and I2V communication. The numerical and simulation results of the cooperative range are calculated based on the distance between infrastructures and vehicles ($d_v$) or targets ($d_T$). 
\begin{figure}[t]
	\centering
	\includegraphics[width=1\linewidth]{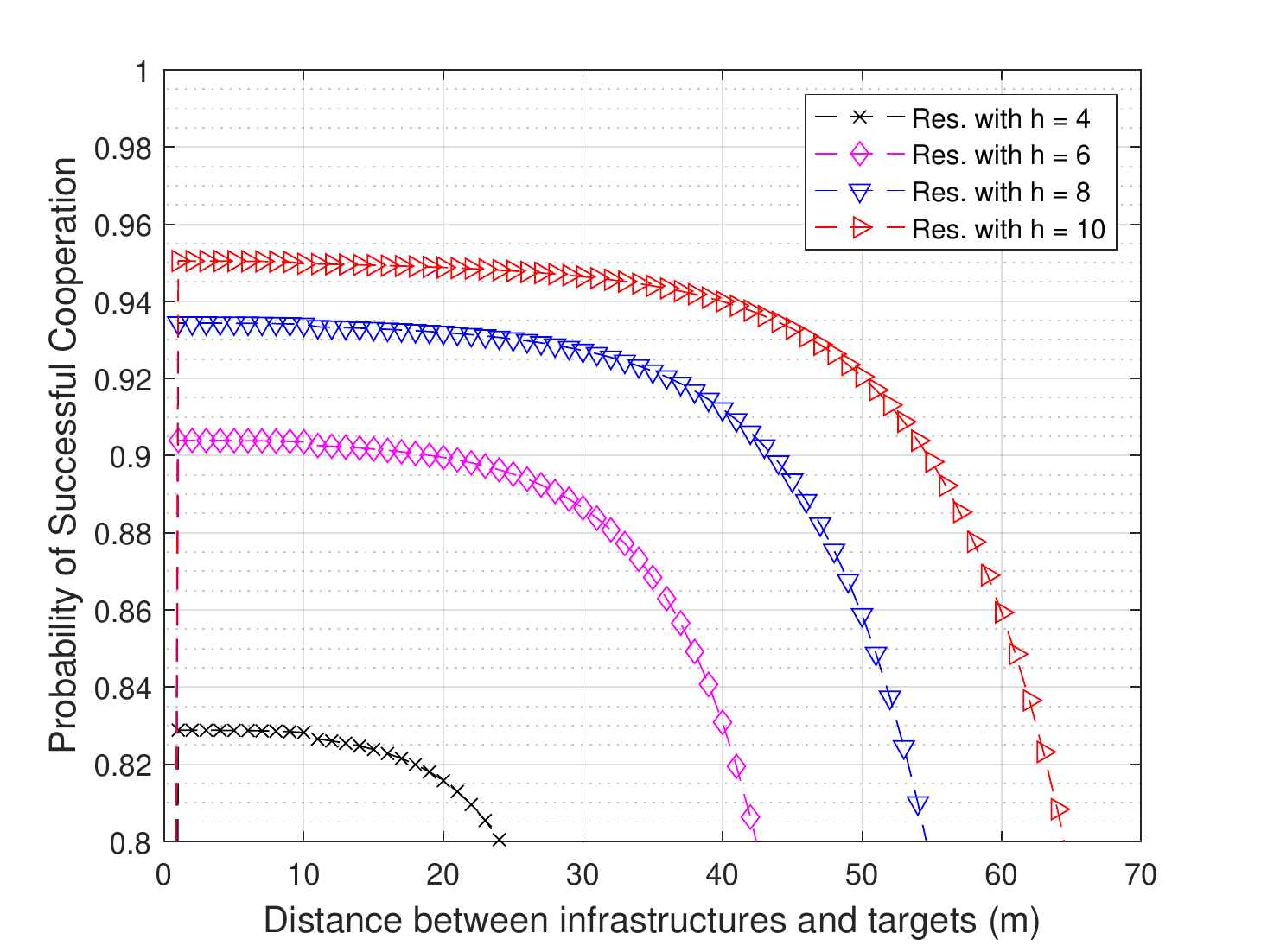}
	\caption{Probability of successful cooperative detection with different detection distance.}
	\label{PrCoopT}
\end{figure}

{Fig. \ref{PrCoopT} shows the results of $\rm Pr_{coop}$ for JCS detection changes with the distance between infrastructures and target obstacles, with I2V cooperation distance being set to 100 m.
As shown in Fig. \ref{PrCoopT}, the initial value of $\rm Pr_{coop}$ is also decided by the infrastructures deploy height. When the height of infrastructure is less than 4 m, the highest $\rm Pr_{coop}$ is less than 80\%, which is also omitted in Fig. \ref{PrCoopT}. }
\begin{figure}[t]
	\centering
	\includegraphics[width=1\linewidth]{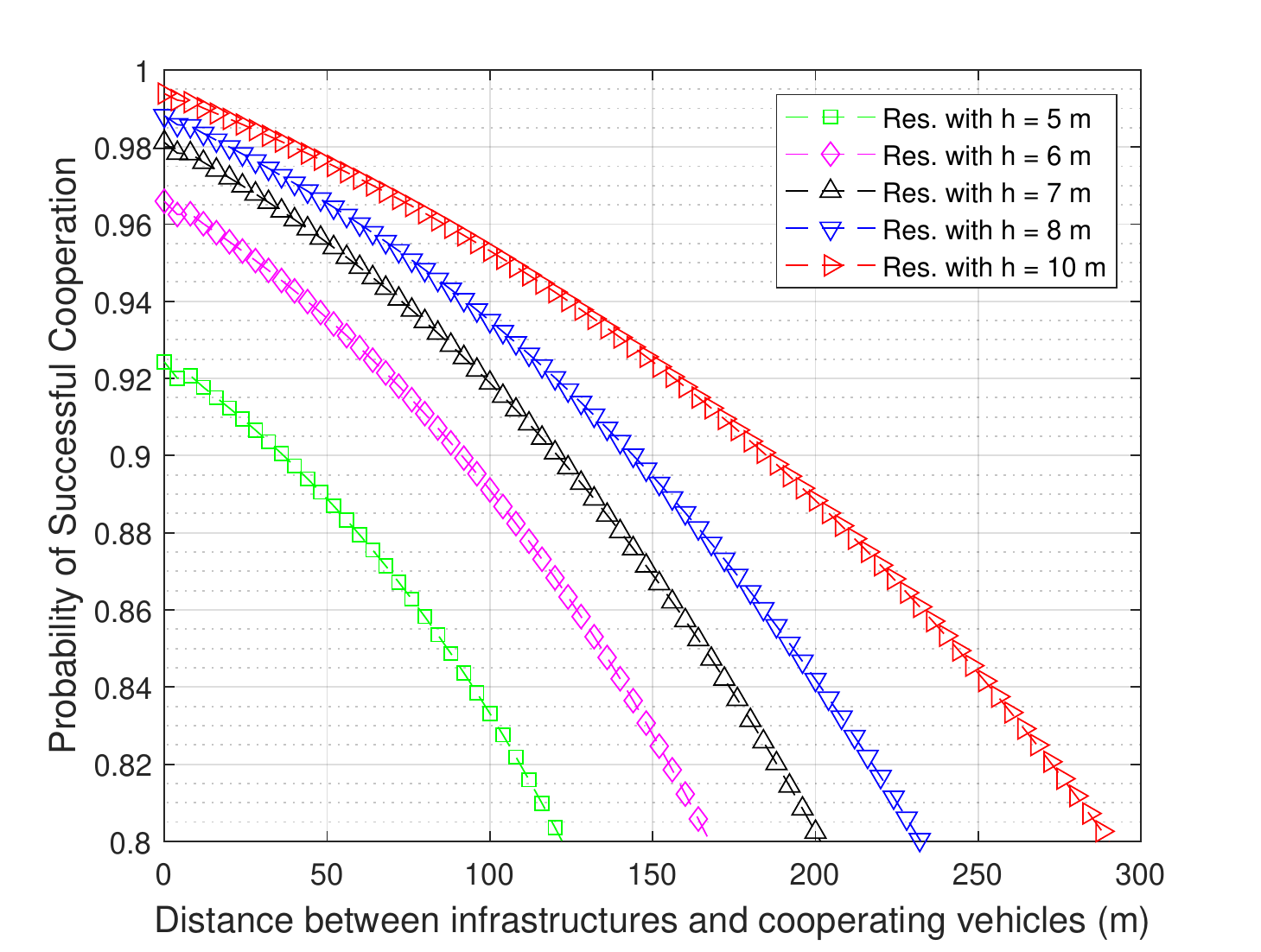}
	\caption{Probability of successful cooperative detection with different communication distance.}
	\label{PrCoopV}
\end{figure}

{Besides the I2O distance $d_T$, the distance between infrastructures and vehicles, known as I2V distance $d_v$, also influences the probability of successful cooperative detection $\rm Pr_{coop}$ as shown in Fig. \ref{PrCoopT}. Since we chose the detection distance of 50 m, JCS detection will not succeed when the height of infrastructures is lower than 5 meters. The effect of successful cooperative detection increase is obvious with the infrastructures height adjusted from 5 m to 7 m. However, the probability of successful cooperative detection is harder to keep the increase rate with further increases in infrastructure height.} 

\begin{figure}[t]
	\centering
	\includegraphics[width=1\linewidth]{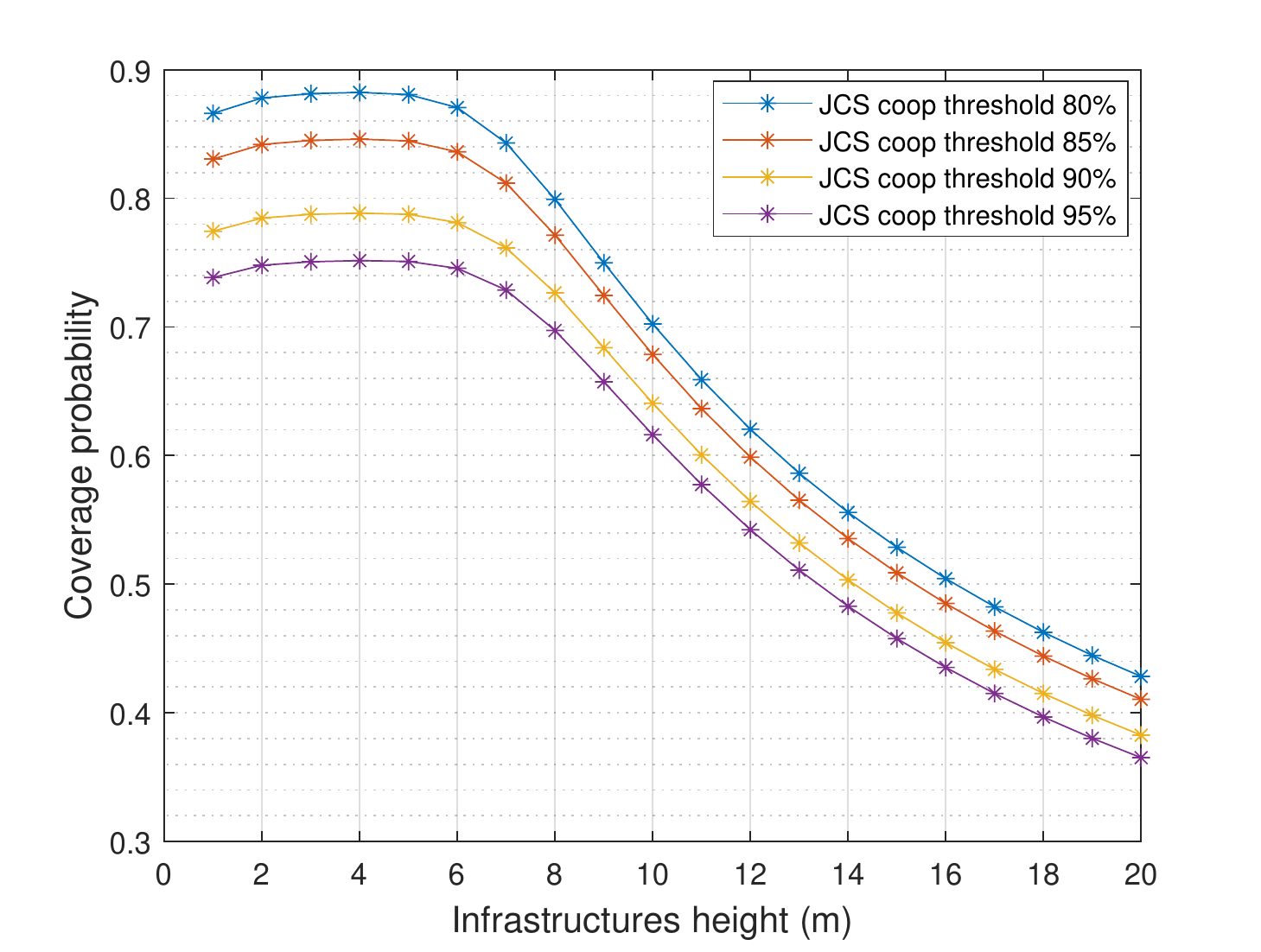}
	\caption{Average detection coverage probability.}
	\label{Rrate}
\end{figure}

{With the result from \textbf{Algorithm 1}, the coverage probability is shown in Fig. \ref{Rrate}.
	As shown in Fig. \ref{Rrate}, the coverage probability slightly increases with lower infrastructures height. As analyzed in Section \ref{SimPrsuccc}, the sparse channels limit the range of communication and JCS detection to a similar value. Thus the coverage probability is relatively stable before lifting the infrastructure's height to 7 m. After 7 m, the increased rate of communication and detection have greater disparity. From our definition of the coverage probability, we set the distance between neighboring infrastructures as the communication range. To fully cover the area between the infrastructures, the JCS detection range should be more than half of the communication range, and due to signal obstruction, there will be another attenuation of the cooperative probability. Thus, further, increasing the deployment height above 7 m will not result in higher coverage performance, as the increase of JCS detection drops fast when the height is over 7 m. The cooperative detection range is then limited by the JCS detection range, and the excess communication range is not useful to the current cooperative detection area but will introduce further interference to neighbor infrastructures.
	Besides, we simulated the coverage probability with the chosen threshold of both JCS detection and I2V communication as 80\%, 85\%, 90\% and 95\%, respectively. It can be concluded that with the higher requirement of probability threshold, the coverage probability will drop, meaning less area is covered due to the limit of performance of the road infrastructures.}
\section{Conclusion}\label{SECV}
In this paper, a novel JCS road infrastructure cooperative detection model is presented with an analysis of the deployment scheme for high-efficiency I2V cooperative detection. Obstacle influence is emphasized to ensure the performance of infrastructures. The GIS statistic of road obstacles and dynamic vehicles are applied to calculate and simulate the process of cooperative detection.{ To analyze the channel sparsity characteristics of the JCS cooperative detection, the LoS transmission probability $\rm Pr_{LoS}$ and SINR for both JCS detection and communication are calculated.} The stochastic geometry approach is applied to analyze the interfering factor to JCS cooperative detection for road infrastructures. Further calculation and simulation of the probability of successful I2V communication and JCS detection are then presented. JCS sensing and communication range is evaluated with the successful detection and communication probability. The probability of successful cooperative detection is then analyzed based on the variant distance of vehicles and targets, and the coverage probability of JCS cooperative detection is finally analyzed. {This paper showed the limited height adjustment range due to the asynchronous change rate of the JCS detection range and communication range and also selected the optimal deploy height of infrastructures at 7 m. The deployment interval of infrastructures is 150 m
	to 300 m according to the requirement of the successful probability threshold, which meets the requirement of China and European transportation standards.}
\appendices

\section{Proof of (\ref{CFS_EPLOS})} \label{Appendix:A}
\setcounter{equation}{0}
\renewcommand{\theequation}{\thesection.\arabic{equation}}

\begin{proof}	\label{proofofEps}
	{
		According to section II, the expectation of received detection signal from line-of-sight channel is calculated as (\ref{EPRS1}). 
		Setting $z = \left[ ({\ln (\frac{x}{d} h-{{\mu}_{0}})})/({\sqrt{2}{{\sigma }_{0}}}) \right]$ and substitute $z$ into $E({{P}_{rs,\rm LoS}})$, we have
		\begin{align}\label{EPSLOS2}
		E({{P}_{rs,\rm LoS}}) &=\int_{\frac{\ln ({h (d-\frac{\pi{r_0}}{2})}/{d}-{\mu_{0}})}{\sqrt{2} {\sigma_{0}}}}^{+\infty }{ {{n}_{p}}{{P}_{res}}x}\frac{{\rm d}{\rm Pr_{Los}}(z)}{{\rm d}x}{\rm d}z\nonumber\\
		&= \frac{1}{h^2} {\sqrt 2 {d^2}\eta {\lambda _0}{n_p}{p_{res}}{r_0}{\sigma _0}} erfc\left[ z \right] \nonumber\\
		&\times \smallint {{\rm{e}}^{2{\mu_0} + 2\sqrt 2 {\sigma _0}z - {\lambda _0}{r_0}\left( { - \frac{{{{\rm{e}}^{ - {z^2}}}}}{{\sqrt \pi }} + z erfc\left[ z \right]} \right)}}{\rm{d}}z,
		\end{align}
		where $erfc$ is the complementary error function. According to series expansion shown in (\ref{Se-Expan-App1}),
		\begin{align}
		e^x = \sum_{n=0}^{\infty}\frac{x^n}{n!},\ 
		erfc(x) = \frac{e^{-x^2}}{x\sqrt{\pi}}\sum_{n = 0}^{\infty}(-1^n)\frac{(2n-1)!}{(2x^2)^n},
		\label{Se-Expan-App1}
		\end{align}
		the integral in (\ref{EPSLOS2}) can be approximated as
		\begin{align}
		&\frac{1}{2} \lambda _0 r_0 {erf}(z)+\frac{\lambda _0 r_0 e^{-z^2}}{\pi }-\frac{\lambda _0 r_0 z^2\left(6 z^2-8 \sqrt{\pi } z+3 \pi \right)}{6 \pi }\nonumber\\
		&-\frac{z^2\left(6 \mu _0+\sqrt{2} \sigma _0\left(4 z-3 \sqrt{\pi }\right)+3\right)}{3 \sqrt{\pi }}+2 \mu _0 z+z.
		\end{align}
		In this way, the expectation of received detection signal from line-of-sight channel can be calculated as
		\begin{align}
		\label{Exp_PLoS_Uz}
		E&({{P}_{rs,\rm LoS}})\nonumber\\
		=& \frac{1}{3\sqrt{2}\pi {{h}^{2}}}{{d}^{2}} {\eta \lambda_0 n_p P_{res}r_0\sigma_0} \left( 3\pi {\lambda_0r_0erf}(z) 
		\right.
		\nonumber\\
		&-\left.
		{\lambda_0r_0}\left( 6{{z}^{2}}-8\sqrt{\pi }z+3\pi \right){{z}^{2}}+6 {\lambda_0r_0}{{e}^{-{{z}^{2}}}}
		\right.
		\nonumber\\
		&-\left.
		2\sqrt{\pi }{{z}^{2}}\left( 6 {\mu_0}+ \sqrt{2} {\sigma_0}\left( 4z-3\sqrt{\pi } \right)+3 \right)
		\right.
		\nonumber\\
		&+\left.
		6\pi (2 {\mu_0}+1)z \right).
		\end{align}
		Applying $z = \left[ \frac{\ln (\frac{x}{d} h-{{\mu}_{0}})}{\sqrt{2}{{\sigma }_{0}}} \right]$ into (\ref{Exp_PLoS_Uz}), the closed form solution of the $E({{P}_{rs,LoS}})$ is shown in (\ref{CFS_EPLOS}).
	}
\end{proof}

\section{Proof of (\ref{Hfunc})} \label{Appendix:B}
\newcounter{mytempeqncnt}
\begin{proof} \label{Theo:{H_{LoS}}}
	{\color{black}
		{\rm 
			{\rm 
				Applying ${\rm {Pr}_{LoS}}(d,{{h}_{1}},{{h}_{2}})$ to $\mathcal{L}_{I_{c,\rm LoS}}$
				\setcounter{equation}{0}, $H_{\rm LoS}$ is then calculated as
				\begin{align}
				&H_{\rm LoS}(\beta,d,h_1,h_2) \nonumber \\
				&= \qquad \frac{\pi }{2 \sqrt{2} \sqrt[4]{\frac{\exp \left(-\frac{1}{2} {\lambda_0} {r_0} ( {\pi} {r_0}-2 d) {erfc}\left(\frac{\log \left(\frac{ {xh}}{d}\right)- {\mu_0}}{\sqrt{2} {\sigma_0}}\right)\right)}{ {\beta_s}}}}.
				\end{align}
				In the same way, the result of $H_{\rm NLoS}$ is
			\begin{align}
				&H_{\rm NLoS}(\beta,d,h_1,h_2) =\nonumber\\
				&\left( -\frac{1}{4} \pi {\beta_s} e^{-d {\lambda_0} {r_0} {erfc}\left(\frac{\log \left(\frac{{xh}}{d}\right)-{\mu_0}} {\sqrt{2} {\sigma_0}}\right)}\right)\nonumber\\
				& \left(e^{d {\lambda_0} {r_0} {erfc}\left(\frac{\log \left(\frac{{xh}}{d}\right)-{\mu_0}}{\sqrt{2} {\sigma_0}}\right)}-e^{\frac{1}{2} {\lambda_0} {pi} {r_0}^2 {erfc}\left(\frac{\log \left(\frac{{xh}}{d}\right)-{\mu_0}}{\sqrt{2} {\sigma_0}}\right)}\right)\nonumber\\
				&/{\left({\beta_s} \left(\exp \left(\frac{1}{2} {\lambda_0} {r_0} ({\pi} {r_0}-2 d) {erfc}\left(\frac{\log \left(\frac{{xh}}{d}\right)-{\mu_0}}{\sqrt{2} {\sigma_0}}\right)\right)-1\right)\right)^{3/4}}
			\end{align}
				}
	}
}
\end{proof}


%


\ifCLASSOPTIONcaptionsoff
 \newpage
\fi


\begin{thebibliography}{10}
	\providecommand{\url}[1]{#1}
	\csname url@samestyle\endcsname
	\providecommand{\newblock}{\relax}
	\providecommand{\bibinfo}[2]{#2}
	\providecommand{\BIBentrySTDinterwordspacing}{\spaceskip=0pt\relax}
	\providecommand{\BIBentryALTinterwordstretchfactor}{4}
	\providecommand{\BIBentryALTinterwordspacing}{\spaceskip=\fontdimen2\font plus
		\BIBentryALTinterwordstretchfactor\fontdimen3\font minus
		\fontdimen4\font\relax}
	\providecommand{\BIBforeignlanguage}[2]{{%
			\expandafter\ifx\csname l@#1\endcsname\relax
			\typeout{** WARNING: IEEEtran.bst: No hyphenation pattern has been}%
			\typeout{** loaded for the language `#1'. Using the pattern for}%
			\typeout{** the default language instead.}%
			\else
			\language=\csname l@#1\endcsname
			\fi
			#2}}
	\providecommand{\BIBdecl}{\relax}
	\BIBdecl
	
	\bibitem{8704212}
	X.~{Yu}, X.~{Chen}, Y.~{Huang}, L.~{Zhang}, J.~{Guan}, and Y.~{He}, ``Radar
	moving target detection in clutter background via adaptive dual-threshold
	sparse fourier transform,'' \emph{IEEE Access}, vol.~7, pp. 58\,200--58\,211,
	May. 2019.
	
	\bibitem{10.1145/3448076}
	Z.~Fang, G.~Wang, X.~Xie, F.~Zhang, and D.~Zhang, ``Urban map inference by
	pervasive vehicular sensing systems with complementary mobility,''
	\emph{Proc. ACM Interact. Mob. Wearable Ubiquitous Technol.}, vol.~5, no.~1,
	Mar. 2021.
	
	\bibitem{9162145}
	N.~Q. Hieu, D.~T. Hoang, N.~C. Luong, and D.~Niyato, ``irdrc: An intelligent
	real-time dual-functional radar-communication system for automotive
	vehicles,'' \emph{IEEE Wireless Communications Letters}, vol.~9, no.~12, pp.
	2140--2143, Aug. 2020.
	
	\bibitem{GBT}
	China-Standard-Administration, ``Taxonomy of driving automation for vehicles,''
	\emph{GB/T 40429-2021.}, Mar.2021.
	
	\bibitem{Intel}
	S.~A. Sehra, \emph{Paving the Way Forward: Intelligent Road
		Infrastructure}.\hskip 1em plus 0.5em minus 0.4em\relax IoT Smart Cities and
	Transportation, Intel Corporation, Aug. 2020.
	
	\bibitem{7934405}
	M.~G. Nilsson, C.~Gustafson, T.~Abbas, and F.~Tufvesson, ``A measurement-based
	multilink shadowing model for v2v network simulations of highway scenarios,''
	\emph{IEEE Transactions on Vehicular Technology}, vol.~66, no.~10, pp.
	8632--8643, May. 2017.
	
	\bibitem{9171304}
	F.~Liu, W.~Yuan, C.~Masouros, and J.~Yuan, ``Radar-assisted predictive
	beamforming for vehicular links: Communication served by sensing,''
	\emph{IEEE Transactions on Wireless Communications}, vol.~19, no.~11, pp.
	7704--7719, Aug. 2020.
	
	\bibitem{9359665}
	X.~Chen, Z.~Feng, Z.~Wei, P.~Zhang, and X.~Yuan, ``Code-division ofdm joint
	communication and sensing system for 6g machine-type communication,''
	\emph{IEEE Internet of Things Journal}, vol.~8, no.~15, pp. 12\,093--12\,105,
	Feb. 2021.
	
	\bibitem{9282206}
	X.~Chen, Z.~Feng, Z.~Wei, F.~Gao, and X.~Yuan, ``Performance of joint
	sensing-communication cooperative sensing uav network,'' \emph{IEEE
		Transactions on Vehicular Technology}, vol.~69, no.~12, pp. 15\,545--15\,556,
	Dec. 2020.
	
	\bibitem{STDGAN}
	``Cd 365 portal and cantilever signs/signals gantries,'' Department for
	Infrastructure, An Roinn Bonneagair, Tech. Rep. CD 365 Revision 1, formerly
	BD 51/14, IAN 193/16, BE 7/04, Mar. 2020.
	
	\bibitem{8108565}
	J.~A. Zhang, A.~Cantoni, X.~Huang, Y.~J. Guo, and R.~W. Heath, ``Joint
	communications and sensing using two steerable analog antenna arrays,'' in
	\emph{2017 IEEE 85th Vehicular Technology Conference (VTC Spring)}, Nov.
	2017, pp. 1--5.
	
	\bibitem{8382292}
	M.~Tahir, S.~S. Afzal, M.~S. Chughtai, and K.~Ali, ``On the accuracy of
	inter-vehicular range measurements using gnss observables in a cooperative
	framework,'' \emph{IEEE Transactions on Intelligent Transportation Systems},
	vol.~20, no.~2, pp. 682--691, Jun. 2019.
	
	\bibitem{8859331}
	M.~N. Sial, Y.~Deng, J.~Ahmed, A.~Nallanathan, and M.~Dohler, ``Stochastic
	geometry modeling of cellular v2x communication over shared channels,''
	\emph{IEEE Transactions on Vehicular Technology}, vol.~68, no.~12, pp.
	11\,873--11\,887, Oct. 2019.
	
	\bibitem{8453027}
	S.~Zhang, H.~Zhang, B.~Di, and L.~Song, ``Joint trajectory and power
	optimization for uav sensing over cellular networks,'' \emph{IEEE
		Communications Letters}, vol.~22, no.~11, pp. 2382--2385, Aug. 2018.
	
	\bibitem{9088261}
	S.~Huang, N.~Jiang, Y.~Gao, W.~Xu, Z.~Feng, and F.~Zhu, ``Radar
	sensing-throughput tradeoff for radar assisted cognitive radio enabled
	vehicular ad-hoc networks,'' \emph{IEEE Transactions on Vehicular
		Technology}, vol.~69, no.~7, pp. 7483--7492, May. 2020.
	
	\bibitem{7506244}
	S.~A.~A. Shah, E.~Ahmed, F.~Xia, A.~Karim, M.~Shiraz, and R.~M. Noor,
	``Adaptive beaconing approaches for vehicular ad hoc networks: A survey,''
	\emph{IEEE Systems Journal}, vol.~12, no.~2, pp. 1263--1277, Jun. 2018.
	
	\bibitem{5776640}
	C.~Sturm and W.~Wiesbeck, ``Waveform design and signal processing aspects for
	fusion of wireless communications and radar sensing,'' \emph{Proceedings of
		the IEEE}, vol.~99, no.~7, pp. 1236--1259, May. 2011.
	
	\bibitem{8461678}
	P.~Kumari, R.~W. Heath, and S.~A. Vorobyov, ``Virtual pulse design for ieee
	802.11ad-based joint communication-radar,'' in \emph{2018 IEEE International
		Conference on Acoustics, Speech and Signal Processing (ICASSP)}, Sept. 2018,
	pp. 3315--3319.
	
	\bibitem{9411464}
	U.~Kumbul, N.~Petrov, F.~van~der Zwan, C.~S. Vaucher, and A.~Yarovoy,
	``Experimental investigation of phase coded fmcw for sensing and
	communications,'' in \emph{2021 15th European Conference on Antennas and
		Propagation (EuCAP)}, Mar. 2021, pp. 1--5.
	
	\bibitem{9285278}
	M.~Bekar, C.~J. Baker, E.~G. Hoare, and M.~Gashinova, ``Joint mimo radar and
	communication system using a psk-lfm waveform with tdm and cdm approaches,''
	\emph{IEEE Sensors Journal}, vol.~21, no.~5, pp. 6115--6124, Mar. 2021.
	
	\bibitem{8304814}
	S.~H. Ahmed, D.~Mu, and D.~Kim, ``Improving bivious relay selection in
	vehicular delay tolerant networks,'' \emph{IEEE Transactions on Intelligent
		Transportation Systems}, vol.~19, no.~3, pp. 987--995, Feb. 2018.
	
	\bibitem{8264740}
	Z.~Zhou, H.~Yu, C.~Xu, Y.~Zhang, S.~Mumtaz, and J.~Rodriguez, ``Dependable
	content distribution in d2d-based cooperative vehicular networks: A big
	data-integrated coalition game approach,'' \emph{IEEE Transactions on
		Intelligent Transportation Systems}, vol.~19, no.~3, pp. 953--964, Jan. 2018.
	
	\bibitem{8302837}
	E.~Ahmed and H.~Gharavi, ``Cooperative vehicular networking: A survey,''
	\emph{IEEE Transactions on Intelligent Transportation Systems}, vol.~19,
	no.~3, pp. 996--1014, Feb. 2018.
	
	\bibitem{8057297}
	J.~Wang, C.~Jiang, K.~Zhang, T.~Q.~S. Quek, Y.~Ren, and L.~Hanzo, ``Vehicular
	sensing networks in a smart city: Principles, technologies and
	applications,'' \emph{IEEE Wireless Communications}, vol.~25, no.~1, pp.
	122--132, 2018.
	
	\bibitem{8265207}
	J.~Wang, C.~Jiang, Z.~Han, Y.~Ren, and L.~Hanzo, ``Internet of vehicles:
	Sensing-aided transportation information collection and diffusion,''
	\emph{IEEE Transactions on Vehicular Technology}, vol.~67, no.~5, pp.
	3813--3825, Jan. 2018.
	
	\bibitem{5948952}
	G.~Karagiannis, O.~Altintas, E.~Ekici, G.~Heijenk, B.~Jarupan, K.~Lin, and
	T.~Weil, ``Vehicular networking: A survey and tutorial on requirements,
	architectures, challenges, standards and solutions,'' \emph{IEEE
		Communications Surveys Tutorials}, vol.~13, no.~4, pp. 584--616, Jul. 2011.
	
	\bibitem{7059538}
	J.-Y. Chang and Y.-W. Chen, ``A cluster-based relay station deployment scheme
	for multi-hop relay networks,'' \emph{Journal of Communications and
		Networks}, vol.~17, no.~1, pp. 84--92, Mar. 2015.
	
	\bibitem{9119440}
	Z.~Fang, Z.~Wei, X.~Chen, H.~Wu, and Z.~Feng, ``Stochastic geometry for
	automotive radar interference with rcs characteristics,'' \emph{IEEE Wireless
		Communications Letters}, vol.~9, no.~11, pp. 1817--1820, Jun. 2020.
	
	\bibitem{8928298}
	A.~S. Kabanov, V.~N. Azarov, and V.~P. Mayboroda, ``An analysis of the use and
	difficulties in introducing information technology and information systems in
	transport and the transport infrastructure,'' in \emph{2019 International
		Conference "Quality Management, Transport and Information Security,
		Information Technologies" (IT QM IS)}, Dec. 2019, pp. 192--196.
	
	\bibitem{TFGAN}
	``Smart mobility and congestion charging,'' The Hong Kong Institution of
	Engineers, LT Division-[Online], Tech. Rep.
	http://www.hkengineer.org.hk/issue/vol48-august2020/cover-story/, Aug. 2020.
	
	\bibitem{ITU-TK20}
	``Recommandation k.20: Resistibility of telecommunication equipment installed
	in a telecommunication centre to overvoltages and overcurrents,''
	Telecommunication Standardization Scrtor, ITU-T, Series K: Production Against
	Interference, Tech. Rep., Jun. 2021.
	
	\bibitem{GIM_cloud}
	``Geographical information monitoring cloud platform,'' - GIM Cloud, [Online],
	Tech. Rep. http://www.dsac.cn/DataProduct/Detail/201801, Dec. 2018.
	
	\bibitem{9052010}
	A.~{Al-Hourani}, ``On the probability of line-of-sight in urban environments,''
	\emph{IEEE Wireless Communications Letters}, vol.~9, no.~8, pp. 1178--1181,
	Mar. 2020.
	
	\bibitem{8855937}
	H.~Ma, Z.~Wei, X.~Chen, Z.~Fang, Y.~Liu, F.~Ning, and Z.~Feng, ``Performance
	analysis of joint radar and communication enabled vehicular ad hoc network,''
	in \emph{2019 IEEE/CIC International Conference on Communications in China
		(ICCC)}, Oct. 2019, pp. 887--892.
	
	\bibitem{9373010}
	R.~He, B.~Ai, G.~Wang, M.~Yang, C.~Huang, and Z.~Zhong, ``Wireless channel
	sparsity: Measurement, analysis, and exploitation in estimation,'' \emph{IEEE
		Wireless Communications}, vol.~28, no.~4, pp. 113--119, Mar. 2021.
	
	\bibitem{8253543}
	W.~{Shi}, H.~{Zhou}, J.~{Li}, W.~{Xu}, N.~{Zhang}, and X.~{Shen}, ``Drone
	assisted vehicular networks: Architecture, challenges and opportunities,''
	\emph{IEEE Network}, vol.~32, no.~3, pp. 130--137, Jan. 2018.
	
	\bibitem{Radar-Handbook}
	M.~I. Skolnik, \emph{Radar Handbook, Third Edition}.\hskip 1em plus 0.5em minus
	0.4em\relax The McGraw-Hill Companies, 2018.
	
	\bibitem{5766057}
	X.~{Huang}, G.~{Wang}, F.~{Hu}, and S.~{Kumar}, ``Stability-capacity-adaptive
	routing for high-mobility multihop cognitive radio networks,'' \emph{IEEE
		Transactions on Vehicular Technology}, vol.~60, no.~6, pp. 2714--2729, May.
	2011.
	
	\bibitem{6043861}
	W.~Wang, T.~Jost, and U.-C. Fiebig, ``Characteristics of the nlos bias for an
	outdoor-to-indoor scenario at 2.45 ghz and 5.2 ghz,'' \emph{IEEE Antennas and
		Wireless Propagation Letters}, vol.~10, pp. 1127--1130, Oct. 2011.
	
	\bibitem{9490660}
	J.~Zheng, S.~Yang, X.~Wang, Y.~Xiao, and T.~Li, ``Background noise filtering
	and clustering with 3d lidar deployed in roadside of urban environments,''
	\emph{IEEE Sensors Journal}, vol.~21, no.~18, pp. 20\,629--20\,639, Jul.
	2021.
	
	\bibitem{8555855}
	Z.~{Wei}, Z.~{Guo}, Z.~{Feng}, J.~{Zhu}, C.~{Zhong}, Q.~{Wu}, and H.~{Wu},
	``Spectrum sharing between uav-based wireless mesh networks and ground
	networks,'' in \emph{2018 10th International Conference on Wireless
		Communications and Signal Processing (WCSP)}, Oct. 2018, pp. 1--6.
	
	\bibitem{6863654}
	A.~{Al-Hourani}, S.~{Kandeepan}, and S.~{Lardner}, ``Optimal lap altitude for
	maximum coverage,'' \emph{IEEE Wireless Communications Letters}, vol.~3,
	no.~6, pp. 569--572, Jul. 2014.
	
	\bibitem{7756327}
	C.~{Zhang} and W.~{Zhang}, ``Spectrum sharing for drone networks,'' \emph{IEEE
		Journal on Selected Areas in Communications}, vol.~35, no.~1, pp. 136--144,
	Jan. 2017.
	
	\bibitem{7412759}
	M.~{Mozaffari}, W.~{Saad}, M.~{Bennis}, and M.~{Debbah}, ``Unmanned aerial
	vehicle with underlaid device-to-device communications: Performance and
	tradeoffs,'' \emph{IEEE Transactions on Wireless Communications}, vol.~15,
	no.~6, pp. 3949--3963, Jun. 2016.
	
	\bibitem{VATNA}
	``Ni mmwave hybrid beamforming testbed reference architecture,'' NATIONAL
	INSTRUMENTS CORP., 11500 North Mopac Expressway Austin, TX
	78759-3504-[Online], Tech. Rep.
	https://semiengineering.com/antenna-array-design-for-adas/, Feb. 2022.
	
	\bibitem{8246333}
	X.~Yang, W.~Lu, N.~Wang, K.~Nieman, C.-K. Wen, C.~Zhang, S.~Jin, X.~Mu,
	I.~Wong, Y.~Huang, and X.~You, ``Design and implementation of a tdd-based
	128-antenna massive mimo prototype system,'' \emph{China Communications},
	vol.~14, no.~12, pp. 162--187, Dec. 2017.
	
	\bibitem{3GPPTR37}
	``Study on evaluation methodology of new vehicle-to-everything v2x use cases
	for lte and nr,'' 3GPP TR 37.885, V15.3.0, Tech. Rep., Jun. 2019.
	
\end{thebibliography}
\end{document}